\documentclass[aps,twocolumn]{revtex4}
\usepackage{graphicx}
\usepackage{epsfig}
\usepackage{dcolumn}
\usepackage{amsmath}
\def\Journal#1#2#3#4{{#1} {\bf#2}, #3 (#4)}

\def\NPA{{\rm Nucl. Phys.} A}
\def\NPB{{\rm Nucl. Phys.} B}
\def\PLB{{\rm Phys. Lett.}  B}

\def\PRD{{\rm Phys. Rev.} D}
\def\PRC{{\rm Phys. Rev.} C}
\def\ZPC{{\rm Z. Phys.} C}
\def\JPG{{\rm J. Phys.} G}
\def\EPJC{{\rm Eur.Phys.J.}C}

\def\ep{\epsilon}
\def\lam{\lambda}

\def\la{\langle}
\def\ra{\rangle}

\def\al{\alpha}

\def\be{\begin{equation}}
\def\ee{\end{equation}}
\def\bea{\begin{eqnarray}}
\def\eea{\end{eqnarray}}
\input{epsfig.sty}
\begin{document}
\title{Semileptonic and radiative decays of the $B_c$ meson in light-front quark model}
\author{ Ho-Meoyng Choi$^{a}$ and Chueng-Ryong Ji$^{b}$\\
$^a$ Department of Physics, Teachers College, Kyungpook National
University,
     Daegu, Korea 702-701\\
$^b$ Department of Physics, North Carolina State University,
Raleigh, NC 27695-8202}
\begin{abstract}
We investigate the exclusive semileptonic $B_c\to
(D,\eta_c,B,B_s)\ell\nu_\ell$, $\eta_b\to
B_c\ell\nu_\ell$($\ell=e,\mu,\tau$) decays
using the light-front quark model constrained by the variational principle
for the QCD motivated effective Hamiltonian. The form factors
$f_+(q^2)$ and $f_-(q^2)$ are obtained from the analytic continuation method in
the $q^+=0$ frame. While the form factor $f_+(q^2)$ is free from the zero-mode,
the form factor $f_-(q^2)$ is not free from the zero-mode in the $q^+=0$
frame. We quantify the zero-mode contributions to $f_-(q^2)$ for various
semileptonic $B_c$ decays. Using our effective method to relate the non-wave function
vertex to the light-front valence wave function, we incorporate the zero-mode
contribution as a convolution of zero-mode operator with the initial and final
state wave functions.
Our results are then compared to the available experimental data and the results
from other theoretical approaches. Since the prediction on the magnetic dipole
$B^*_c\to B_c+\gamma$ decay turns out to be very sensitive to the mass difference between
$B^*_c$ and $B_c$ mesons, the decay width $\Gamma(B^*_c \to B_c \gamma)$ may help in
determining the mass of $B^*_c$ experimentally.
Furthermore, we compare the results from the harmonic oscillator potential and the linear
potential and identify the decay processes that are sensitive to the choice of
confining potential. From the future experimental data on these sensitive processes, one
may obtain more realistic information on the potential between quark and antiquark in the
heavy meson system.
\end{abstract}


\maketitle
\section{Introduction}
The exclusive semileptonic decay processes of heavy mesons generated a
great excitement not only in extracting the most accurate values
of Cabbibo-Kobayashi-Maskawa(CKM) matrix elements but also in
testing diverse theoretical approaches to describe the internal
structure of hadrons. The great virtue of semileptonic decay processes is
that the effects of the strong interaction can be separated from
the effects of the weak interaction into a set of
Lorentz-invariant form factors, i.e., the essential informations of
the strongly interacting quark/gluon structure inside hadrons.
Thus, the theoretical problem associated with analyzing
semileptonic decay processes is essentially that of calculating the weak
form factors.

In particular, along with the experimental study planned both at
the Tevatron and at the Large Hadron Collider(LHC), the study of
the $B_c$ meson has been very interesting due to its
outstanding feature; i.e., the $B_c$ meson is the lowest bound state of
two heavy ($b,c$) quarks with different flavors.  Because of the
fact that the $B_c$ meson carries the flavor explicitly, not like
the symmetric heavy quarkonium ($b\bar{b}$, $c\bar{c}$) states, there
is no gluon or photon annihilation via strong interaction or
electromagnetic interaction. It can decay only via weak
interaction. Since both $b$-and $c$-quarks forming the $B_c$ meson
are heavy, the $B_c$ meson can decay appreciably not only through
the $b\to q$($q=c,u$) transition with $c$ quark being a spectator
but also through the $c\to q$($q=s,d$) transition with $b$ quark being
a spectator. The former transitions correspond to the semileptonic
decays to $\eta_c$ and $D$ mesons, while the latter transitions correspond
to the decays to $B_s$ and $B$ mesons. The latter transitions
are governed typically by much larger CKM matrix element; e.g.,
$|V_{cs}|\sim 1$ for $B_c\to B_s\ell\nu_\ell(\ell=e,\mu)$, vs.
$|V_{cb}|\sim 0.04$ for $B_c\to \eta_c\ell\nu_\ell(\ell=e,\mu,\tau)$.
For this reason, although the phase space in $c\to s,d$
transitions is much smaller than that in $b\to c,u$ transitions,
the $c$-quark decays provide about $\sim 70\%$ to the decay width
of $B_c$. The $b$-quark decays and weak annihilation add
about
20$\%$
and 10$\%$, respectively~\cite{Gouz}. This indicates that
both $b$-and $c$-quark decay processes contribute on a comparable
footing to the $B_c$ decay width.

There are many theoretical approaches to the calculation of
exclusive $B_c$ semileptonic decay modes. Although we may not be able
to list them all, we may note here the following works: QCD sum
rules~\cite{Gouz, CNP,KKL,HZ}, the relativistic quark
model~\cite{IKS01,IKS05,IKS06} based on an effective Lagrangian
describing the coupling of hadrons to their constituent quarks,
the quasipotential approach to the relativistic quark
model~\cite{EFG67,EFG03D,EFG03E}, the instantaneous
nonrelativistic approach to the Bethe-Salpeter(BS)
equation~\cite{CC94}, the relativistic quark model based on the BS
equation~\cite{LC97,AMV}, the QCD relativistic potential
model~\cite{CD00}, the relativistic quark-meson model~\cite{NW},
the nonrelativistic quark model~\cite{HNV}, the covariant
light-front quark model~\cite{WSL}, and the constituent quark
model~\cite{LM,DW,DSV,God} using BSW(Bauer, Stech, and Wirbel)
model~\cite{BSW} and ISGW(Isgur, Scora, Grinstein, and Wise)
model~\cite{ISGW}.

The purpose of this paper is to extend our light-front
quark model(LFQM)~\cite{CJ1,CJ2,JC,CJK02,Choi07,Choi08}  based on the
QCD-motivated effective LF Hamiltonian to calculate the hadronic form
factors and decay widths for the exclusive semileptonic $B_c\to P
\ell\nu_{\ell}(P=D,D_s,B,B_s)$ and $\eta_b\to B_c \ell\nu_{\ell}$ decays
and the magnetic dipole $B^*_c\to B_c\gamma$ transition.
In our previous LFQM
analysis~\cite{CJ1,CJ2,JC,CJK02,Choi07,Choi08}, we have analyzed
the meson mass spectra~\cite{CJ1,CJ2} and various exclusive processes of
the ground state pseudoscalar($P$) and vector($V$) mesons such as
the $P\to P$ semileptonic heavy/light meson decays~\cite{CJ2,JC}, the rare
$B\to K\ell^+\ell^-$ decays~\cite{CJK02}, and the
magnetic dipole transitions
of
the low-lying heavy/light
pseudoscalar/vector mesons~\cite{CJ1,Choi07,Choi08}.  In those analyses,
we found a good agreement with the experimental data.
However, since we didn't analyze the $B_c$ and $B^*_c$ mesons yet,
we shall extend our LFQM to predict the masses and the decay constants of
$B_c$ and $B^*_c$ mesons as well as the above mentioned exclusive
decays of $B_c$ and $B^*_c$ mesons.

Our LFQM~\cite{CJ1,CJ2,JC,CJK02,Choi07,Choi08}
analysis in this work has several salient features:
(1) We have implemented the
variational principle to the QCD motivated effective LF
Hamiltonian to enable us to analyze the meson mass spectra and to
find optimized model parameters. The present investigation
further constrains the phenomenological parameters and extends
the applicability of our LFQM to the wider range of hadronic phenomena.
(2) We have performed the analytical continuation from the
spacelike region to the physical timelike region to obtain the
weak form factor $f_+(q^2)$ for the exclusive semileptonic decays
between the two pseudoscalar mesons as well as to obtain the decay form
factors $F_{VP}(q^2)$ for $V\to P\gamma^*$ transitions. The
Drell-Yan-West($q^+=q^0+q^3=0$) frame (i.e., $q^2=-{\bf
q}^2_\perp<0$) is useful because only the valence contributions are needed
unless the zero-mode contribution exists.

The form factor $f_+(q^2)$ can be obtained just from the valence contribution in the
$q^{+}= 0$ frame without encountering the zero-mode contribution~\cite{Zero}.
However, the form factor $f_-(q^2)$ receives the higher Fock state contribution(i.e.,
the zero-mode in the $q^+=0$ frame or the nonvalence contribution
in the $q^+> 0$ frame) within the framework of LF quantization. Thus, it
is necessary to include either the zero-mode contribution(if
working in the $q^+=0$ frame) or the nonvalence contribution(if
working in the $q^+> 0$ frame) to obtain the form factor $f_-(q^2)$.
In this work, we utilize our effective method presented in~\cite{JC}
to express the zero-mode contribution as a convolution of zero-mode operator
that we find in this work with the initial and final state LF wave functions.
In this way, we calculate the form factor $f_-(q^2)$ in the $q^+=0$ frame
with the perpendicular components of the currents and discuss
the LF covariance of $f_-(q^2)$ in the valence region by analyzing
the covariant BS model and the LF covariant analysis
described by Jaus~\cite{Jaus99}. We also estimate the zero-mode contributions
to the $f_-(q^2)$ for various semileptonic $B_c$ decays in our LFQM.

The paper is organized as follows. In Sec. II,
we discuss the $P\to P$ semileptonic decays using an exactly
solvable model based on the covariant BS model of $(3+1)$-dimensional
fermion field theory. We explicitly show the equivalence between the
results obtained by
the manifestly covariant method and the LF method in the $q^+=0$ frame.
The extraction of the zero-mode contribution to
$f_-(q^2)$ in the $q^+=0$ frame and the effective inclusion of the zero-mode
in the valence region are discussed.
In Sec. III, we  briefly
describe the formulation of our LFQM and the procedure of fixing
the model parameters using the variational principle for the QCD
motivated effective Hamiltonian. The masses and decay constants of
the $B^*_c$ and $B_c$ mesons are predicted and compared with the
data as well as other theoretical model predictions. The
distribution amplitudes(DAs) for the heavy-flavored mesons
such as $D,\eta_c,B,B_s,B_c$ and $\eta_b$ are also
obtained in this section. In Sec. IV, we calculate the weak form
factors $f_+(q^2)$ and $f_-(q^2)$ in the $q^+=0$ frame using the
plus and perpendicular components of the currents, respectively.
The zero-mode contribution to the form factor $f_-(q^2)$ is also discussed.
In Sec. V, the decay form factor
$F_{B^*_c B_c}(q^2)$ for the $B^*_c\to B_c\gamma^*$ transition and
the decay width for $B^*_c\to B_c\gamma$ are presented. The coupling
constant $g_{B^*_c B_c}$ needed for the calculation of the decay
width for $B^*_c\to B_c\gamma$ is determined in the limit
$q^2\to 0$, i.e., $g_{B^*_c B_c}=F_{B^*_c B_c}(q^2=0)$.
For the numerical calculation of the semileptonic and radiative decays,
the form factors $f_{\pm}(q^2)$ for the
semileptonic decays and $F_{B^*_c B_c}(q^2)$ for the $B^*_c\to B_c\gamma^*$
transition are analytically continued to the timelike $q^2>0$
region by changing ${\bf q}^2_\perp$ to $-q^2$ in the form factor.
In Sec. VI, our numerical results (i.e., the form factors and decay rates for
$B_c\to(D,\eta_c,B,B_s)\ell\nu_\ell$, $\eta_b\to B_c\ell\nu_\ell$,
and $B^*_c\to B_c\gamma^{(*)}$ decays) are presented and compared
with the experimental data as well as other theoretical results.
Summary and discussion follow in Sec.VII.

\section{$P\to P$ semileptonic decays in covariant Bethe-Salpeter model}

\subsection{Manifestly covariant calculation}
The amplitude $A$ for a semileptonic decay of a meson
$Q_{1}\bar{q}$ with the four-momentum $P_1$ and the mass $M_1$ into
another meson $Q_{2}\bar{q}$ with the four-momentum $P_2$ and the mass $M_2$
is given by
\be\label{SA} A=\frac{G_F}{\sqrt{2}}V_{Q_1\bar{Q}_2}L_\mu H^\mu,
\ee where $G_{F}$ is the Fermi constant, $V_{Q_{1}\bar{Q}_{2}}$ is
the relevant CKM mixing matrix element, $L_\mu$ is the lepton
current
 \be\label{Lmu} L_\mu
=\bar{u}_{\nu_\ell}\gamma_\mu(1-\gamma^5)v_\ell,
 \ee
 and $H^\mu$ is the hadron current \be\label{Hmu} H^\mu=\la
P_2,\ep|(V^\mu-A^\mu)|P_1\ra. \ee Here, $\ep$ is the polarization
of the daughter meson and $V^\mu$ and $A^\mu$ are the vector and
axial vector currents, respectively. If the final state is
pseudoscalar, the hadron current can be decomposed as follows:
\bea\label{PPF}
\la P_2|A^\mu|P_1\ra &=&0, \nonumber\\
\la P_2|V^\mu|P_{1}\ra&=& f_{+}(q^2)(P_1 + P_2)^{\mu} +
f_-(q^2)q^\mu,
 \eea
 where $q^\mu=(P_1-P_2)^\mu$ is the
four-momentum transfer to the lepton pair($\ell\nu_\ell$) and
$m^2_\ell\leq q^2\leq (M_1-M_2)^2$. Sometimes it is useful to
express the matrix element of the vector current in terms of
$f_+(q^2)$ and $f_0(q^2)$, which correspond to the transition
amplitudes with $1^-$ and $0^+$ spin-parity quantum numbers in the
center of mass of the lepton pair, respectively. They satisfy the
following relation: \be f_0(q^2) = f_+(q^2) +
\frac{q^2}{M^2_1-M^2_2}f_-(q^2). \ee

Including the nonzero lepton mass, the differential decay rate for the
exclusive $0^{-}\to 0^{-}\ell\nu_\ell$ process is given by\cite{CJK}
\bea\label{Drate} \frac{d\Gamma}{dq^{2}}&=&
\frac{G^{2}_{F}}{24\pi^{3}} |V_{Q_{1}\bar{Q}_{2}}|^{2}
K(q^{2})\biggl(1-\frac{m^2_\ell}{q^2}\biggr)^2 \nonumber\\
&&\times\biggl\{[K(q^2)]^2\biggl(1+\frac{m^2_\ell}{2q^2}\biggr)
|f_{+}(q^{2})|^{2}\nonumber\\
&&+
M^2_1\biggl(1-\frac{M^2_2}{M^2_1}\biggr)^2\frac{3}{8}
\frac{m^2_l}{q^2}|f_0(q^2)|^2\biggr\},
\eea where $K(q^{2})$ is the kinematic factor given by
\be\label{Kin} K(q^{2})= \frac{1}{2M_{1}}\sqrt{
(M_{1}^{2}+M_{2}^{2}-q^{2})^{2}-4M_{1}^{2}M_{2}^{2}}. \ee

  The solvable model, based on the covariant Bethe-Salpeter(BS) model of
($3+1$)-dimensional fermion field theory~\cite{BCJ01,BCJ03,MF}, enables us to derive the transition form
factors between two pseudoscalar mesons explicitly. The matrix element
${\cal M}^\mu\equiv\la P_2|V^\mu|P_1\ra$ in this BS model is given by
 \be\label{ap:3}
 {\cal M}^\mu = ig_1 g_2
\Lambda^2_1\Lambda^2_2\int\frac{d^4k}{(2\pi)^4} \frac{S^\mu} {N_{\Lambda_1}
N_{1} N_{\bar q} N_{2} N_{\Lambda_2}},
 \ee
where $g_1$ and $g_2$ are the normalization factors which can be fixed by
requiring both charge form factors of pseudoscalar mesons to be unity at zero
momentum transfer, respectively.  To regularize the covariant fermion
triangle-loop in ($3+1$) dimensions, we replace the point gauge-boson vertex
$\gamma^\mu (1-\gamma_5)$ by a non-local (smeared) gauge-boson vertex
$({\Lambda_1}^2 / N_{\Lambda_1})\gamma^\mu (1-\gamma_5)
( {\Lambda_2}^2 / N_{\Lambda_2})$, where $N_{\Lambda_1}
=p_1^2-{\Lambda_1}^2+i\ep$ and $N_{\Lambda_2}
=p_2^2-{\Lambda_2}^2+i\ep$, and thus the factor
$({\Lambda_1}{\Lambda_2})^2$ appears in the normalization factor. $\Lambda_1$
and $\Lambda_2$ play the role of momentum cut-offs similar to the Pauli-Villars
regularization~\cite{BCJ01,BCJ03}.  The rest of the denominators in Eq.~(\ref{ap:3}),
i.e., $N_{1} N_{\bar q} N_{2}$, are coming from the intermediate fermion
propagators in the triangle loop diagram and are given by
\bea\label{ap:4}
N_{1} &=& p_1^2 -{m_1}^2 + i\ep, \nonumber \\
N_{\bar q} &=& k^2 - m^2_{\bar q} + i\ep, \nonumber \\
N_{2} &=& p_2^2 -{m_2}^2 + i\ep,
 \eea
where $m_1$, $m_{\bar q}$, and  $m_2$ are the masses of the constituents carrying the intermediate
four-momenta $p_1=P_1 -k$, $k$, and $p_2=P_2 -k$, respectively.
Furthermore, the trace term in Eq.~(\ref{ap:3}), $S^\mu$, is given by
 \bea\label{ap:5}
 S^\mu &=& {\rm Tr}[\gamma_5(\not\!p_1 + m_1)\gamma^\mu (\not\!p_2
+m_2)\gamma_5(-\not\!k + m_{\bar q})]
 \nonumber\\
 &=&4 (k\cdot P_2 - k^2 + m_2m_{\bar q}) P^\mu_1
 \nonumber\\
 &&+ 4(k\cdot P_1 - k^2 + m_1m_{\bar q}) P^\mu_2
 \nonumber\\
 &&+ 4(k^2 - P_1\cdot P_2 - m_1m_{\bar q} - m_2m_{\bar q} + m_1m_2)k^\mu.
 \nonumber\\
 \eea
We then decompose the product of five denominators given in Eq.~(\ref{ap:3}) as
follows:
\bea\label{ap:6}
  \frac{1}{N_{\Lambda_1} N_{1} N_{\bar q} N_{2} N_{\Lambda_2}}
&=& \frac{1}{({\Lambda_1}^2-{m_1}^2)({\Lambda_2}^2-{m_2}^2)}
\nonumber\\
&&\times \frac{1}{N_{\bar q}}\biggl( \frac{1}{N_{\Lambda_1}} -
\frac{1}{N_{}}\biggr) \biggl(\frac{1}{N_{\Lambda_2}} - \frac{1}{N_{2}} \biggr).
\nonumber\\
\eea

Once we reduce the five propagators into a sum of terms containing
three propagators using Eq.~(\ref{ap:6}), we use the Feynman
parametrization for the three propagators, e.g.,
%
 \bea\label{ap:7}
\frac{1}{N_{1} N_{\bar q} N_{2}} &=&
\int^1_0 dx \int^{1-x}_0  dy
\nonumber\\
&&\times
\frac{2}{[N_{\bar q} + (N_1  -  N_{\bar q}) x + (N_2 - N_{\bar q}) y]^3}.
\nonumber\\
 \eea
%
We then make a Wick rotation of Eq.~(\ref{ap:3}) in $D$-dimensions to
regularize the integral, since otherwise one looses the logarithmically
divergent terms in Eq.~(\ref{ap:3}).  Following the above procedure, we
finally obtain the Lorentz-invariant form factors  $f_+(q^2)$ and $f_-(q^2)$ as
follows:
\begin{widetext}
\bea\label{ap:8}
  f_+(q^2) &=& \frac{N}{8\pi^2 ({\Lambda_1}^2-{m_1}^2)({\Lambda_2}^2-{m_2}^2) }\int^1_0 dx\int^{1-x}_0 dy
\biggl\{ [ 3(x+y) -4]\ln\biggl(\frac{C_{\Lambda_1 m_2}C_{m_1 \Lambda_2}}
{C_{\Lambda_1 \Lambda_2}C_{m_1 m_2}} \biggr)
 \nonumber\\
 &&+ \biggl [ (1-x-y)^2 (x M^2_1 + y M^2_2)
 + x y(2 - x -y) q^2
 + (x + y) (m_1 m_2 - m_1m_{\bar q} - m_2 m_{\bar q}) + m_{\bar q}(m_1 + m_2)\biggr ] C
 \biggr\}, \nonumber \\
f_-(q^2) &=& \frac{ N}{8\pi^2
({\Lambda_1}^2-{m_1}^2)({\Lambda_2}^2-{m_2}^2)}\int^1_0 dx\int^{1-x}_0 dy
 \biggl\{ 3(x-y)
 \ln \biggl(\frac{C_{\Lambda_1 m_2}C_{m_1 \Lambda_2}}
{C_{\Lambda_1 \Lambda_2}C_{m_1 m_2}} \biggr) +  \biggl[  (y M^2_2 - x M^2_1)
 \nonumber\\
 && +  (x^2-y^2) (x M^2_1 + y M^2_2)
  - x y (x-y) q^2 + (x-y)(m_1m_2 - m_1m_{\bar q} - m_2m_{\bar q}) + m_{\bar q}(m_2 - m_1) \biggr]C \biggr\},
\nonumber \\
\eea
where $N= g_1 g_2\Lambda^2_1\Lambda^2_2$ and
 $C=(1/C_{\Lambda_1 \Lambda_2} -1/C_{\Lambda_1 m_2} - 1/C_{m_1 \Lambda_2} +
1/C_{m_1 m_2})$ with
 \bea\label{ap:9}
 C_{\Lambda_1 \Lambda_2}&=& (1-x-y)(x M_1^2 + y
M_2^2) + xy q^2- (x \Lambda_1^2 + y \Lambda_2^2) - (1-x-y)m^2_{\bar q}, \nonumber \\
C_{\Lambda_1 m_2}&=& (1-x-y)(x M_1^2 + y M_2^2) + xy q^2
- (x \Lambda_1^2 + y m_2^2) - (1-x-y)m^2_{\bar q}, \nonumber \\
C_{m_1 \Lambda_2}&=& (1-x-y)(x M_1^2 + y M_2^2) + xy q^2
- (x m_1^2 + y \Lambda_2^2) - (1-x-y)m^2_{\bar q}, \nonumber \\
C_{m_1 m_2}&=& (1-x-y)(x M_1^2 + y M_2^2) + xy q^2
- (x m_1^2 + y m_2^2) - (1-x-y)m^2_{\bar q}.
 \eea
%
Note that the logarithmic terms in $f_+(q^2)$ and $f_-(q^2)$ are obtained from the
dimensional regularization with the Wick rotation.

\subsection{Light-front calculation}
Performing the LF calculation of Eq.~(\ref{ap:3}) in the $q^+=0$ frame in parallel with
the manifestly covariant calculation, we shall use the plus and
perpendicular components of the currents to obtain the form factors $f_+(q^2)$ and
$f_-(q^2)$, respectively. That is, in the $q^+=0$ frame, one obtains the relations between the current matrix
elements and the weak form factors as follows
 \bea\label{ap:17}
 f_+(q^2) &=& \frac{{\cal M}^+}{2P^+_1},
\nonumber\\
 f_-(q^2) &=& f_+(q^2) + \frac{ {\cal M}^\perp \cdot {\bf q}_\perp}{ {\bf q}^2_\perp}.
 \eea

The LF calculation for the trace term in
Eq.~(\ref{ap:5})  can be separated into the on-shell propagating part
$S^{\mu}_{\rm on}$ and the instantaneous part $S^{\mu}_{\rm inst}$ via
 \be\label{ap:10}
 \not\!p + m = (\not\!p_{\rm on} + m) + \frac{1}{2}\gamma^+(p^- - p^-_{\rm on})
 \ee
 as
 \be\label{ap:11}
   S^\mu  = S^\mu_{\rm on} + S^\mu_{\rm inst},
 \ee
where
 \bea\label{ap:12}
  S^\mu_{\rm on} &=&
4 \biggl[
 p^\mu_{1\rm on} (p_{2\rm on}\cdot k_{\rm on}) - k^\mu_{\rm on} (p_{1\rm on}\cdot p_{2\rm on})
+ p^\mu_{2\rm on} (p_{1\rm on}\cdot k_{\rm on})
  + m_2m_{\bar q} p^\mu_{1\rm on} + m_1m_{\bar q}p^\mu_{2\rm on}
 + m_1m_2 k^\mu_{\rm on} \biggr],
 \eea
 and
  \bea\label{ap:13}
  S^\mu_{\rm inst} &=&
  2(p^-_1 - p^-_{1\rm on}) \biggl[ p^\mu_{2\rm on} k^+_{\rm on} - p^+_{2\rm on} k^\mu_{\rm on}
   + g^{\mu +} (p_{2\rm on}\cdot k_{\rm on} + m_2 m_{\bar q}) \biggr]
   \nonumber\\
&& + 2(p^-_2 - p^-_{2\rm on}) \biggl [ p^\mu_{1\rm on} k^+_{\rm on} - p^+_{1\rm on} k^\mu_{\rm on}
  + g^{\mu +} (p_{1\rm on}\cdot k_{\rm on} + m_1 m_{\bar q}) \biggr]
  \nonumber\\
  &&+ 2( k^- - k^-_{\rm on})  \biggl[ p^\mu_{1\rm on} p^+_{2\rm on} + p^+_{1\rm on} p^\mu_{2\rm on}
   - g^{\mu +} (p_{1\rm on}\cdot p_{2\rm on} - m_1 m_2) \biggr]
  \nonumber\\
  &&+ 2g^{\mu+}k^+_{\rm on} (p^-_1 - p^-_{1\rm on}) (p^-_2 - p^-_{2\rm on}).
 \eea
%
%
 Note that the subscript (on) denotes the on-mass-shell ($p^2=m^2$) quark momentum,
i.e., $p^-=p^-_{\rm on} = (m^2+{\bf p}^2_\perp)/p^+$. The traces in Eqs.
(\ref{ap:12}) and (\ref{ap:13}) are then obtained as
 \bea\label{ap:14}
  S^+_{\rm on} &=& \frac{4 P^+_1}{1-x} ({\bf k}_\perp\cdot{\bf k'}_\perp
  + {\cal A}_1{\cal A}_2 ),
\nonumber\\
 S^+_{\rm inst} &=& 0,
  \eea
  for the plus component of the currents and
 \bea\label{ap:15}
  S^\perp_{\rm on}
&=&\frac{-2 {\bf k}_\perp}{x (1-x)} \biggl[
 2{\bf k}_\perp\cdot{\bf k'}_\perp + (1-x) ({\bf q}^2_\perp + m^2_1 + m^2_2)
 + 2x^2 m^2_{\bar q} + 2x (1-x) (m_1m_{\bar q} + m_2m_{\bar q} - m_1m_2) \biggr]
 \nonumber\\
 &&-\frac{2{\bf q}_\perp}{x (1-x)} ( {\bf k}^2_\perp + {\cal A}^2_1),
 \nonumber\\
 S^\perp_{\rm inst} &=& -2P^+_1 \biggl[ (p^-_1 - p^-_{1\rm on}){\bf k'}_\perp
+ (p^-_2 - p^-_{2\rm on}) {\bf k}_\perp
 + x(k^- - k^-_{\rm on}) (2{\bf k}_\perp + {\bf q}_\perp) \biggr],
 \eea
for the perpendicular components of the currents, where
${\bf k'}_\perp = {\bf k}_\perp + (1-x){\bf q}_\perp$ and
${\cal A}_i  =  (1-x) m_i + x m_{\bar q}\; (i=1,2)$.

As one can see from Eqs.~(\ref{ap:14}) and (\ref{ap:15}),
the perpendicular components of the currents receive instantaneous contributions
while the plus component of the currents does not receive them.
Especially, the absence of the instantaneous contributions to the plus current
indicates that there is no zero-mode contribution to the hadronic matrix element
of the plus current.

\subsubsection{Valence contribution}
In the valence region $0<k^+<P^+_2$, the pole $k^-=k^-_{\rm on}=({\bf
k}^2_\perp + m^2_{\bar q} -i\ep)/k^+$ (i.e., the spectator quark) is located in the lower half of
the complex $k^-$-plane.  Thus, the Cauchy integration formula for the
$k^-$ integral in Eq.~(\ref{ap:3}) gives
%
  \bea\label{ap:18}
 {\cal M}^\mu_{ val}&=&
\frac{N}{16\pi^3}\int^1_0 \frac{dx}{(1-x)}\int d^2{\bf k}_\perp
\chi_1(x,{\bf k}_\perp) \chi_2 (x, {\bf k'}_\perp)
 S^\mu_{val} ,
  \eea
%
where $S^\mu_{val} =S^\mu_{\rm on} + S^\mu_{\rm inst}(k^-=k^-_{\rm on})$ and
 \be\label{ap:19}
M^2_0 = \frac{{\bf k}^2_\perp + m^2_{\bar q}}{1-x} + \frac{{\bf k}^2_\perp + m^2_1}{x},
\;\;
M'^2_0 = \frac{{\bf k'}^2_\perp + m^2_{\bar q}}{1-x} + \frac{{\bf k'}^2_\perp + m^2_2}{x},
 \ee
and $M^2_{\Lambda_1}=M^2_0(m_1\to\Lambda_1)$, $M'^2_{\Lambda_2}=
M'^2_0(m_2\to\Lambda_2)$ with ${\bf k'}_\perp={\bf k}_\perp + (1-x) {\bf
q}_\perp$.
The LF vertex functions $\chi_1$ and $\chi_2$ are given by
 \be\label{ap:21}
 \chi_1(x,{\bf k}_\perp) = \frac{1}{ x^2 (M^2_1 - M^2_0)(M^2_1-M^2_{\Lambda_1})},
 \;\;
\chi_2 (x, {\bf k'}_\perp)= \frac{1}{ x^2 (M^2_2 -
M'^2_0)(M^2_2-M'^2_{\Lambda_2})}.
\ee
From Eqs.~(\ref{ap:14}) and (\ref{ap:15}), we obtain the valence
contribution to ${\cal M}^+_{val}$  and ${\cal M}^\perp_{val}$ as follows
 \bea\label{ap:20}
 {\cal M}^+_{val}&=& \frac{NP^+_1}{4\pi^3}\int^1_0 \frac{dx}{(1-x)^2}
 \int d^2{\bf k}_\perp \chi_1(x,{\bf k}_\perp)
 \chi_2 (x, {\bf k'}_\perp)
 ( {\bf k}_\perp\cdot{\bf k'}_\perp + {\cal A}_1{\cal A}_2 ),
  \nonumber\\
 {\cal M}^\perp_{val}&=& \frac{N}{16\pi^3}\int^1_0 \frac{dx}{(1-x)}
 \int d^2{\bf k}_\perp \chi_1(x,{\bf k}_\perp)
 \chi_2 (x, {\bf k'}_\perp) S^\perp_{val},
  \eea
where
 \be\label{ap:22}
  S^\perp_{val} = -2{\bf k}_\perp \biggl[ M^2_1 + M^2_2 + {\bf q}^2_\perp -(m_1 -m_{\bar q})^2
  - (m_2 - m_{\bar q})^2 + (m_1 - m_2)^2 \biggr]
  -2 {\bf q}_\perp \biggl[ (1-x) M^2_1 + xM^2_0 - (m_1-m_{\bar q})^2 \biggr].
  \ee
From Eqs.~(\ref{ap:17}) and~(\ref{ap:20}), we get the LF valence contributions to $f_+(q^2)$
 and $f_-(q^2)$ as follows
 %
 \bea\label{ap:23}
f^{val}_+(q^2)&=& \frac{N}{8\pi^3}\int^1_0 \frac{dx}{(1-x)^2}
 \int d^2{\bf k}_\perp \chi_1 (x, {\bf k}_\perp)
  \chi_2(x, {\bf k'}_\perp)
   ( {\bf k}_\perp\cdot{\bf k'}_\perp + {\cal A}_1{\cal A}_2 ),
 \nonumber\\
 f^{val}_-(q^2) &=& \frac{N}{8\pi^3}\int^1_0 \frac{dx}{(1-x)}
 \int d^2{\bf k}_\perp \chi_1 (x, {\bf k}_\perp)
 \chi_2(x, {\bf k'}_\perp)
 \nonumber\\
 &&\times
  \biggl\{
  - (1-x) M^2_1 + (m_2 - m_{\bar q}){\cal A}_1 - m_{\bar q}(m_1 - m_{\bar q})
   + \frac{ {\bf k}_\perp\cdot{\bf q}_\perp }{q^2} [ M^2_1 + M^2_2
  - 2(m_1 - m_{\bar q})(m_2 - m_{\bar q}) ]
  \biggr\}.
  \nonumber\\
  \eea
 We note that the form factors in Eq.~(\ref{ap:23}) obtained in the spacelike region using
 the $q^+=0$ frame are analytically continued to the timelike region by changing ${\bf q}^2_\perp$
 to $-q^2$ in the form factors.

\subsubsection{Zero-mode contribution }
In the nonvalence region $P^+_2<k^+<P^+_1$, the poles are at $p^-_1=p^-_{1\rm
on}(m_1) = [m^2_1 +{\bf k}^2_\perp -i\ep]/p^+_1$ (from the struck quark propagator)
and $p^-_1=p^-_{1\rm on}(\Lambda_1) =  [\Lambda^2_1+{\bf k}^2_\perp -i\ep]/
p^+_1$ (from the smeared quark-photon vertex), which are located in the upper
half of the complex $k^-$-plane. In order to estimate the zero-mode contribution,
we define $\al=P^+_2/P^+_1=1- q^+/P^+_1=1-\beta$ and then the region
$P^+_2<k^+<P^+_1$ corresponds to $\al< 1-x < 1$ or equivalently $0<x<\beta$.
That is, the zero-mode contribution($q^+\to 0$) to the hadronic matrix element is
obtained from the $\beta\to 0$(i.e. $x\to 0$) limit for the integration of  the
longitudinal momentum $x$. The fact that $S^+_{\rm on}$ in Eq.~(\ref{ap:14}) is
regular in the $x\to 0$ limit implies no zero-mode contribution to $f_+(q^2)$.
However, as one can see from Eq.~(\ref{ap:15}), both $S^\perp_{\rm on}$ and
$S^\perp_{\rm inst}$  include the terms proportional to $1/x ({\rm i.e.}\; p^-_1)$, which
are singular as $x\to 0$. Those singular terms in the perpendicular current may
be the source of zero-mode contribution to the hadronic matrix element in the
nonvalence region.

When we do the Cauchy integration over $k^-$ to obtain the LF time-ordered
diagrams, we use Eq.~(\ref{ap:6}) to avoid the complexity of treating double
$p^-_1$-poles. As mentioned above, the zero-mode contribution comes from the
$p^-_i(i=1,2)$ factors in $S^\perp_{nv}=S^\perp_{\rm on} +S^\perp_{\rm inst}$. For
instance, we define the zero-mode contribution to the  $1/(N_{\bar q} N_{\Lambda_1}
N_{\Lambda_2})$ term in Eq.~(\ref{ap:6}) having $p^-_1=p^-_{1\rm
on}(\Lambda_1)$ pole as
\be\label{ap:24}
 [ {\cal M}^\perp_{\Lambda_1\Lambda_2}]_{\rm Z.M.}
  =i N \lim_{\beta\to 0}\int_{nv}\frac{d^4k}{(2\pi)^4}
 \frac{S^\perp_{nv}(p^-_1=p^-_{1\rm on}(\Lambda_1)) }
 {N_{\bar q} N_{\Lambda_1}N_{\Lambda_2}}.
 \ee
The zero-mode contributions to the other three terms in Eq.~(\ref{ap:6}) can be
defined the same way as in Eq.~(\ref{ap:24}) to give the net zero-mode
contribution ${\cal M}^\perp_{\rm Z.M.} = [ {\cal
M}^\perp_{\Lambda_1\Lambda_2}(p^-_{1\rm on}(\Lambda_1)) ]_{\rm Z.M.} - [ {\cal
M}^\perp_{\Lambda_1 m_2}(p^-_{1\rm on}(\Lambda_1)) ]_{\rm Z.M.} - [ {\cal
M}^\perp_{m_1\Lambda_2}(p^-_{1\rm on}(m_1)) ]_{\rm Z.M.} + [ {\cal M}^\perp_{m_1
m_2}(p^-_{1\rm on}(m_1))]_{\rm Z.M.}$. Essentially, the nonvanishing zero-mode
contributions in Eq.~(\ref{ap:24}) are summarized as follows:

 \bea\label{ap:25}
  (1)&& Z = p^-_1, p^-_2, -k^-  \;{\rm in}\; S^\perp_{nv}:
 \nonumber\\
 &&\lim_{\beta\to 0}\int_{nv}\frac{d^4k}{(2\pi)^4}
 \frac{Z}{N_{\bar q} N_{\Lambda_1}N_{\Lambda_2}}
 =\frac{i}{16\pi^3} \int^1_0 dz\int d^2{\bf k}_\perp
 \frac{ \Lambda^2_{1\perp} }
 { \Lambda^2_{1\perp} [z \Lambda^2_{2\perp} + (1 -z)\Lambda^2_{1\perp}]},
 \nonumber\\
 (2)&& Z= p^-_{1\rm on} \;{\rm in}\; S^\perp_{nv}:
\nonumber\\
 &&\lim_{\beta\to 0}\int_{nv}\frac{d^4k}{(2\pi)^4}
 \frac{Z}{N_{\bar q} N_{\Lambda_1}N_{\Lambda_2}}
 =\frac{i}{16\pi^3} \int^1_0 dz\int d^2{\bf k}_\perp
 \frac{ m^2_{1\perp} }
 { \Lambda^2_{1\perp} [z \Lambda^2_{2\perp} + (1 -z)\Lambda^2_{1\perp}]},
 \nonumber\\
 (3)&& Z= p^-_{2\rm on} \;{\rm in}\; S^\perp_{nv}:
\nonumber\\
 &&\lim_{\beta\to 0}\int_{nv}\frac{d^4k}{(2\pi)^4}
 \frac{Z}{N_{\bar q} N_{\Lambda_1}N_{\Lambda_2}}
 =\frac{i}{16\pi^3} \int^1_0 dz\int d^2{\bf k}_\perp
 \frac{ m^2_{2\perp} z/(z-1)}
 { \Lambda^2_{1\perp} [z \Lambda^2_{2\perp} + (1 -z)\Lambda^2_{1\perp}]},
 \eea
where the variable change $x =\beta z$ was made and $m^2_{i\perp} = m^2_i +
{\bf p}^2_{i\perp}$ and $\Lambda^2_{i\perp}=\Lambda^2_i + {\bf p}^2_{i\perp}$. We
should note that $Z=p^-_{i\rm on}k^+ \to p^-_{i\rm on}P^+_1(i=1,2)$  in
$S^\perp_{nv}$ and other terms such as  $Z=p^-_{1\rm on}p^+_2$
and
$p^-_{2\rm on}p^+_1$
go
to zero in the $\beta\to 0$ limit. Finally, we get
the following nonvanishing zero-mode contribution to the trace term
$S^\perp_{nv}$ in Eq.~(\ref{ap:24}):
 \be\label{ap:26}
S^\perp_{\rm Z.M.}=\lim_{x\to 0}S^\perp_{nv}
 = 2 p^-_1({\bf p}_{1\perp}+{\bf p}_{2\perp})=2 p^-_1(2{\bf p}_{1\perp}-{\bf q}_{\perp}),
 \ee
which in fact is common to other three terms in Eq.~(\ref{ap:6}). After a little
manipulation, we finally get the following nonvanishing zero-mode contribution to
the form factor $f_-(q^2)$ in Eq.~(\ref{ap:17}) as follows
 \bea\label{ap:27}
 f^{\rm Z.M.}_-(q^2)&=& i\frac{N}{({\Lambda_1}^2-{m_1}^2)({\Lambda_2}^2-{m_2}^2)}
 \int \frac{d^4k}{(2\pi)^4} \frac{ S^\perp_{\rm Z.M.} \cdot {\bf q}_\perp} { {\bf
q}^2_\perp}
 \biggl( \frac{1}{N_{\bar q}N_{\Lambda_1}N_{\Lambda_2}}
 - \frac{1}{N_{\bar q} N_{\Lambda_1} N_2}
 - \frac{1}{N_{\bar q} N_1 N_{\Lambda_2}}
 + \frac{1}{N_{\bar q} N_1 N_2} \biggr)
 \nonumber\\
 &=& \frac{N}{8\pi^2({\Lambda_1}^2-{m_1}^2)({\Lambda_2}^2-{m_2}^2)}
 \int^1_0 dz (1-2z) \ln\biggl( \frac{B_{\Lambda_1 m_2}
 B_{m_1\Lambda_2}} {B_{\Lambda_1\Lambda_2}B_{m_1m_2}} \biggr),
  \nonumber\\
 \eea
where
 \bea\label{ap:28}
B_{\Lambda_1 \Lambda_2} &=& z(1-z) {\bf q}^2_\perp + (1-z)\Lambda^2_1 + z
\Lambda^2_2,
 \nonumber\\
B_{\Lambda_1 m_2} &=& z(1-z) {\bf q}^2_\perp + (1-z)\Lambda^2_1 + z m^2_2,
 \nonumber\\
B_{m_1 \Lambda_2} &=& z(1-z) {\bf q}^2_\perp + (1-z)m^2_1 + z \Lambda^2_2,
 \nonumber\\
 B_{m_1 m_2} &=& z(1-z) {\bf q}^2_\perp + (1-z)m^2_1 + z m^2_2.
 \eea
 Therefore, we get the LF covariant weak form factors in the $q^+=0$ frame as
 $f^{\rm LFCov}_+(q^2)=f^{val}_+(q^2)$ and
 $f^{\rm LFCov}_-(q^2)=f^{val}_-(q^2) + f^{\rm Z.M.}_-(q^2)$.

\subsubsection{Effective inclusion of the zero-mode in the valence region }
In this exactly solvable covariant BS model, we find that while the matrix element
of the plus current is exactly on-mass shell physical amplitude, that of the perpendicular
current is the off-mass shell amplitude.
As shown in our previous work~\cite{JC}, we can relate the non-wave-function
vertex to the ordinary valence wave function in the $q^+>0$ frame using the iteration of the
irreducible kernel involved in the bound state equation. In the $q^+\to 0$ frame, the
nonvalence contribution in the $q^+>0$ frame corresponds to the zero-mode contribution
in the $q^+=0$ frame. Thus, we can identify the zero-mode operator that is convoluted with
the initial and final state valence wave functions to generate the zero-mode contribution.
Our method can also be realized effectively by the method presented by Jaus~\cite{Jaus99}
using the orientation of the light-front plane characterized by the invariant equation
$\omega\cdot x=0$~\cite{CDKM,SCC}, where $\omega$ is an arbitrary light-like
four vector. The special case $\omega=(1,0,0,-1)$ corresponds to the light-front or null plane
$\omega\cdot x=x^+=0$. While the exact on-shell amplitudes(such as ${\cal M}^+$) should not depend
on the orientation of the light-front plane, the off-shell matrix elements(such as ${\cal M}^\perp$)
acquire a spurious $\omega$ dependence. This problem is closely associated with the violation of
rotational invariance in the computation of the matrix element of a one-body current.
In order to treat the complete Lorentz structure of a hadronic matrix
element the authors in~\cite{CDKM,Jaus99} have developed a method to identify and separate spurious
contributions and to determine the physical, i.e. $\omega$ independent contributions to the hadronic
form factors. Below, we summarize the result of zero-mode contribution obtained
from the method by Jaus~\cite{Jaus99} and discuss the equivalence with
our result of zero-mode contribution.

By adopting the $\omega$ dependent light-front covariant approach as in~\cite{CDKM,Jaus99},
we derive the light-front covariant form of the form factor $f_-(q^2)$, which effectively includes
the zero-mode contribution in the valence region. In order to do this, we first decompose the four vector
$p^\mu_1$ in terms of $P=(P_1+P_2), q$, and $\omega$ with $\omega=(1,0,0,-1)$ as follows~\cite{Jaus99}
 \be\label{eq:j1}
p^\mu_1 = P^\mu A^{(1)}_1 + q^\mu A^{(1)}_2 + \frac{1}{\omega\cdot P}\omega^\mu C^{(1)}_1.
 \ee
The coefficients in Eq.~(\ref{eq:j1}) are given by
 \bea\label{eq:j2}
A^{(1)}_1&=& \frac{\omega\cdot p_1}{\omega\cdot P}=\frac{x}{2},
\nonumber\\
A^{(1)}_2&=& \frac{1}{q^2}\biggl( p_1\cdot q
- (q\cdot P)\frac{\omega\cdot p_1}{\omega\cdot P}\biggr)
=\frac{x}{2} + \frac{{\bf k_\perp\cdot{\bf q}_\perp}}{q^2},
\nonumber\\
C^{(1)}_1 &=& p_1\cdot P - P^2 A^{(1)}_1 - q\cdot P A^{(1)}_2 = Z_2 - N_{\bar q},
 \eea
where $N_{\bar q}=k^2-m^2_{\bar q}$ and
 \be\label{eq:j3}
Z_2 = x(M^2_1 - M^2_0) + m^2_1 - m^2_{\bar q} + (1-2x)M^2_1
- [q^2 + q\cdot P]\frac{{\bf k_\perp\cdot{\bf q}_\perp}}{q^2}.
 \ee
Note that only the coefficient $C^{(1)}_1$ which is combined with $\omega^\mu$
depends on $p^-_1$(i.e. zero-mode). In this exactly solvable BS model,
the zero-mode contribution from $p^-_1$ is exactly opposite to that from $N_{\bar q}$,
i.e.
 \bea\label{eq:j4}
 I[p^-_1]_{Z.M.}&=&i\int_{Z.M.} \frac{d^4k}{(2\pi)^4}
\frac{p^-_1}{N_{\Lambda_1}N_1 N_{\bar q} N_2 N_{\Lambda_2}}
\nonumber\\
&=& \frac{N}{16\pi^2({\Lambda_1}^2-{m_1}^2)({\Lambda_2}^2-{m_2}^2)}
\int^1_0 dz  \ln\biggl( \frac{B_{\Lambda_1 m_2}
 B_{m_1\Lambda_2}} {B_{\Lambda_1\Lambda_2}B_{m_1m_2}} \biggr)
 \nonumber\\
 &=&-I[N_{\bar q}]_{Z.M.}.
 \eea
Furthermore, the zero-mode contribution $I[N_{\bar q}]_{Z.M.}$ from $N_{\bar q}$ is exactly the
same as the valence contribution $I[Z_2]_{val}$ from $Z_2$, where $I[Z_2]_{val}$ is given by
 \be\label{eq:j5}
I[N_{\bar q}]_{Z.M.}= I[Z_2]_{val}= \frac{1}{16\pi^3}\int^1_0 \frac{dx}{1-x}\int d^2{\bf k}_\perp
 \chi_1(x,{\bf k}_\perp)\chi_2(x,{\bf k}'_\perp) Z_2.
 \ee

From the identities in Eqs.~(\ref{eq:j4}) and (\ref{eq:j5}), the replacement
$N_{\bar q}\to Z_2$(or equivalently $p^-_1\to -Z_2$) in the spurious $\omega$
dependent (i.e. the zero-mode related) term
$C^{(1)}_1$ in Eq.~(\ref{eq:j2}) becomes covariant, i.e. free
from any $\omega$ dependence.
Effectively, the zero-mode contribution from $p^-_1$ in the valence region
can be given by Eq.~(\ref{eq:j5}).
Using this,
we can effectively include
the zero-mode contribution from the second term $p^-_1{\bf q}_\perp$ in Eq.~(\ref{ap:26})
in the valence region. On the other hand, since the first term $p^-_1{\bf p}_{1\perp}$
in Eq.~(\ref{ap:26}) has a tensor structure, we need the tensor decomposition~\cite{Jaus99}
 \bea\label{eq:j6}
p^\mu_1p^\nu_1 &=& g^{\mu\nu}A^{(2)}_1 + P^\mu P^\nu A^{(2)}_2
+ (P^\mu q^\nu + q^\mu P^\nu)A^{(2)}_3 + q^\mu q^\nu A^{(2)}_4
+ \frac{1}{\omega\cdot P}(P^\mu\omega^\nu + \omega^\mu P^\nu)B^{(2)}_1
\nonumber\\
&&+ \frac{1}{\omega\cdot P}(q^\mu\omega^\nu + \omega^\mu q^\nu)C^{(2)}_1
+ \frac{1}{(\omega\cdot P)^2}\omega^\mu\omega^\nu C^{(2)}_2,
 \eea
 where
 \bea\label{eq:j7}
 A^{(2)}_1 &=& -{\bf k}^2_\perp - \frac{ ({\bf k}_\perp\cdot{\bf q}_\perp)^2}{q^2}, \;\;
 A^{(2)}_2 = [A^{(1)}_1]^2, \;\; A^{(2)}_3 = A^{(1)}_1 A^{(1)}_2,\;\;
 A^{(2)}_4 = [A^{(1)}_2]^2 - \frac{1}{q^2}A^{(2)}_1,
 \nonumber\\
 B^{(2)}_1 &=& A^{(1)}_1 C^{(1)}_1 - A^{(2)}_1,\;\;
 C^{(2)}_1 = A^{(1)}_2 C^{(1)}_1 + \frac{q\cdot P}{q^2} A^{(2)}_1,\;\;
 C^{(2)}_2 = [C^{(1)}_1]^2 + \biggl[ P^2 - \frac{(q\cdot P)^2}{q^2}\biggr] A^{(2)}_1.
 \eea
We note that the coefficients $A^{(2)}_i(i=1,\cdots,4)$ are related with $\mu,\nu=+$ or $\perp$ components and
$B^{(2)}_1$ with $(\mu,\nu)=(+,-)$. According to our power counting rules mentioned above, those
terms are zero-mode free.
On the other hand, the coefficients $C^{(2)}_1$
and $C^{(2)}_2$ are related with  $(\mu,\nu)=(-,\perp)$ and $(-,-)$, respectively. That is, the
$C$ terms are related with the zero-mode contributions. Specifically,  $C^{(2)}_1$
and $C^{(2)}_2$ are related with the zero-mode contribution to the perpendicular and minus
components of the currents, respectively.
The zero-mode contribution from $p^-_1{\bf p}_{1\perp}$ is thus related with $C^{(2)}_1$ and
the effective inclusion of the zero-mode in the valence region can be achieved by setting  $C^{(2)}_1=0$.
This leads to
\be\label{eq:j8}
A^{(1)}_2N_{\bar q} \to A^{(1)}_2Z_2
+ \frac{q\cdot P}{q^2}A^{(2)}_1\;\;
{\rm or}\;\;
p^-_1{\bf p}_{1\perp} \to -{\bf q}_\perp\biggl[ A^{(1)}_2Z_2
+ \frac{q\cdot P}{q^2}A^{(2)}_1\biggr].
\ee
In summary, the zero-mode contribution from $S^\perp_{Z.M.}$ given by Eq.~(\ref{ap:26}) can
be expressed in terms of the zero-mode operator convoluted with the initial and final state LF
vertex functions:
\be\label{eq:j9}
I[S^\perp]_{Z.M.}=\frac{1}{16\pi^3}\int^1_0\frac{dx}{(1-x)}\int d^2{\bf k}_\perp
\chi_1(x,{\bf k}_\perp)\chi_2(x,{\bf k}'_\perp)\biggl\{
-4{\bf q}_\perp\biggl[ A^{(1)}_2Z_2
+ \frac{q\cdot P}{q^2}A^{(2)}_1\biggr] + 2{\bf q}_\perp Z_2 \biggr\},
 \ee
as expected from our effective method presented in our previous work~\cite{JC}.
Consequently, the LF covariant form of the form factor $f_-(q^2)$ is obtained as
 \bea\label{ap:29}
 f^{\rm LFCov}_-(q^2) &=& \frac{N}{8\pi^3} \int^1_0 \frac{dx}{(1-x)}
 \int d^2{\bf k}_\perp \chi_1 (x, {\bf k}_\perp) \chi_2(x, {\bf k'}_\perp) \biggl\{
 -x(1-x) M^2_1 - {\bf k}^2_\perp - m_1m_{\bar q} + (m_2 - m_{\bar q}){\cal A}_1
 \nonumber\\
 && + 2\frac{q\cdot P}{q^2} \biggl[ {\bf k}^2_\perp
 + 2\frac{ ( {\bf k}_\perp \cdot {\bf q}_\perp)^2 } {q^2} \biggr]
 + 2 \frac{ ( {\bf k}_\perp \cdot {\bf q}_\perp)^2 } {q^2}
  + \frac{ {\bf k}_\perp \cdot {\bf q}_\perp } {q^2}  [ M^2_2 - (1-x) (q^2 + q\cdot P) + 2 x M^2_0
  \nonumber\\
  && - (1 - 2x) M^2_1 - 2(m_1 - m_{\bar q}) (m_1 + m_2) ] \biggr\},
 \eea
\end{widetext}
 where $q\cdot P = M^2_1 - M^2_2$.

\begin{figure}
\vspace{0.8cm}
\includegraphics[width=3in,height=3in]{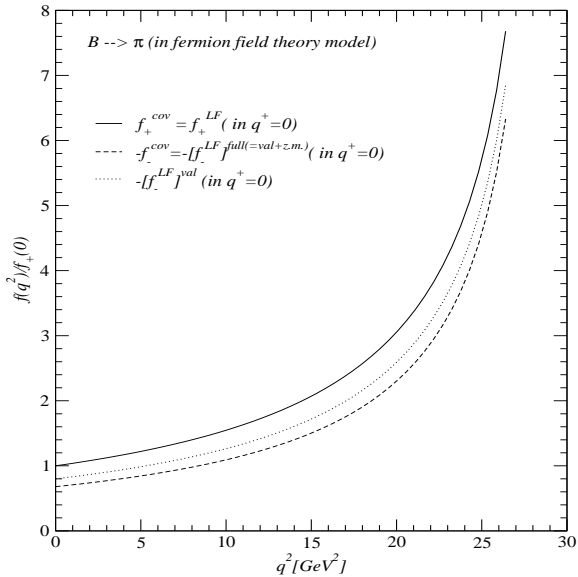}
\caption{The normalized weak form factors $f_{\pm}(q^2)/f_+(0)$ for $B\to \pi$
semileptonic decays obtained from the exactly solvable covariant BS model of fermion
field theory.}
\label{fig4}
\end{figure}

In Fig.~\ref{fig4}, we show our results of the normalized weak
form factors $f_{\pm}(q^2)/f_+(0)$ for the semileptonic $B\to \pi$ decay obtained from
the exactly solvable covariant BS model of fermion field theory.
The used model parameters for $B$ and $\pi$ mesons are $M_B=5.28$ GeV,
$M_\pi=0.14$ GeV, $m_b=4.9$ GeV, $m_{u(d)}=0.43$
GeV, $\Lambda_b=10$ GeV, and $\Lambda_{u(d)}=1.5$ GeV.
These parameters are fixed from the normalization conditions of the $\pi$ and
$B$ elastic form factors at $q^2=0$~\cite{BCJ03}.

The solid line represents the form factors $f_+(q^2)/f_+(0)$ obtained from the
manifestly covariant result [Eq.~(\ref{ap:8})] and
from the full LF calculation $f^{\rm LFCov}_+(q^2)=f^{val}_+(q^2)$ [Eq.~(\ref{ap:23})]
in the $q^+=0$ frame.
Since the two results are in complete agreement with each other, we depict them by the single
solid line.
The dashed line represents the form factors $-f_-(q^2)/f_+(0)$ obtained from the
manifestly covariant result [Eq.~(\ref{ap:8})] and
from the full LF calculation
$f^{\rm LFCov}_-(q^2)=f^{val}_-(q^2) + f^{\rm Z.M.}_-(q^2)$ [Eq.~(\ref{ap:29})]
in the $q^+=0$ frame. Here again, the two results are in complete agreement with each other.
The dotted line represents only the valence contribution  to
$-f_-(q^2)/f_+(0)$.
The difference between the dashed and dotted lines amounts to the zero-mode contribution to the
form factor $f_-(q^2)$.
Although our result for the $f^{\rm LFCov}_-(q^2)$ is essentially
the same as that obtained from Jaus~\cite{Jaus99},
the distinguished features of our approach in deriving the LF covariant
form factor may be summarized as follows:
(1) We separate the trace term into the on-shell propagating part $S^\mu_{\rm on}$
and the instantaneous part $S^\mu_{\rm inst.}$ which enable us to classify the
on-shell and off-shell matrix elements explicitly.
From this one can easily find which matrix element depends on the orientation of
the light-front plane or equivalently receives zero-mode contributions.
(2) Our power counting rule for $p^-_1$ is very efficient method in identifying the
zero-mode terms such as $p^-_1$ and $p^-_1{\bf p}_{1\perp}$ that appear in the
perpendicular currents and $p^-_1p^-_1$ appearing in the minus current.
(3) We explicitly show that the $\omega$ dependent(i.e. zero-mode related) coefficients
$C^{(1)}_1, C^{(2)}_1$, and $C^{(2)}_2$ correspond to $p^-_1,p^-_1{\bf p}_{1\perp}$
and $p^-_1p^-_1$ terms, respectively. These features in our approach should be distinguished
from the approach presented in~\cite{Jaus99}.

While the manifestly covariant BS model of fermion field theory model is good for
the qualitative analysis of semileptonic decays, it is still
semi-realistic. We thus discuss
more phenomenological LFQM and the LF covariant form factors within our LFQM in the
following sections.

\section{Model Description}
The key idea in our LFQM~\cite{CJ1,CJ2} for mesons is to treat the
radial wave function as a trial function for the variational
principle to the QCD-motivated effective Hamiltonian saturating
the Fock state expansion by the constituent quark and antiquark.
The QCD-motivated Hamiltonian for a description of the ground
state meson mass spectra is given by
\bea\label{Ham}
H_{q\bar{q}}|\Psi^{JJ_z}_{nlm}\ra&=&\biggl[ \sqrt{m^2_q+{\vec
k}^2}+\sqrt{m^2_{\bar{q}}+{\vec k}^2}+V_{q\bar{q}}\biggr]
|\Psi^{JJ_z}_{nlm}\ra,
\nonumber\\
&=&[H_0 + V_{q\bar{q}}]|\Psi^{JJ_z}_{nlm}\ra
=M_{q\bar{q}}|\Psi^{JJ_z}_{nlm}\ra, \eea where ${\vec k}=({\bf
k}_\perp, k_z)$ is the three-momentum of the constituent quark,
$M_{q\bar{q}}$ is the mass of the meson, and
$|\Psi^{JJ_z}_{nlm}\ra$ is the meson wave function. In this work,
we use two interaction potentials $V_{q\bar{q}}$; (1) Coulomb
plus harmonic oscillator(HO) and (2) Coulomb plus linear confining
potentials. The hyperfine interaction essential to
distinguish pseudoscalar($0^{-+}$) and vector($1^{--}$) mesons
is also included; viz.,
\be\label{pot} V_{q\bar{q}}=V_0 + V_{\rm hyp} = a + {\cal V}_{\rm
conf}-\frac{4\al_s}{3r} +\frac{2}{3}\frac{{\bf S}_q\cdot{\bf
S}_{\bar{q}}}{m_qm_{\bar{q}}} \nabla^2V_{\rm coul}, \ee where
${\cal V}_{\rm conf}=br(r^2)$ for the linear (HO) potential and
$\la{\bf S}_q\cdot{\bf S}_{\bar{q}}\ra=1/4(-3/4)$ for the vector
(pseudoscalar) meson.
Using this Hamiltonian, we analyze the meson mass spectra and
various wave-function-related observables, such as decay
constants, electromagnetic form factors of mesons in a spacelike
region, and the weak form factors for the exclusive semileptonic
and rare decays of pseudoscalar mesons in the timelike
region~\cite{CJ1,CJ2,JC,CJK02,Choi07,Choi08}.

The momentum-space light-front wave function of the ground state
pseudoscalar and vector mesons is given by \be\label{w.f}
\Psi^{JJ_z}_{100}(x_i,{\bf k}_{i\perp},\lam_i) ={\cal
R}^{JJ_z}_{\lam_1\lam_2}(x_i,{\bf k}_{i\perp}) \phi(x_i,{\bf
k}_{i\perp}), \ee where $\phi(x_i,{\bf k}_{i\perp})$ is the radial
wave function and ${\cal R}^{JJ_z}_{\lam_1\lam_2}$ is the
spin-orbit wave function that is obtained by the interaction-
independent Melosh transformation from the ordinary
spin-orbit wave function assigned by the quantum numbers
$J^{PC}$. The model wave function in Eq.~(\ref{w.f}) is
represented by the Lorentz-invariant internal variables, $x_i=p^+_i/P^+$,
${\bf k}_{i\perp}={\bf p}_{i\perp}-x_i{\bf P}_\perp$ and $\lam_i$,
where $P^\mu=(P^+,P^-,{\bf P}_\perp) =(P^0+P^3,(M^2+{\bf
P}^2_\perp)/P^+,{\bf P}_\perp)$ is the momentum of the meson $M$,
and $p^\mu_i$ and $\lam_i$ are the momenta and the helicities of
constituent quarks, respectively.

The covariant forms of the spin-orbit wave functions
for pseudoscalar and vector mesons are given by
\bea\label{R00_A}
{\cal R}_{\lam_1\lam_2}^{00}
&=&\frac{-\bar{u}_{\lam_1}(p_1)\gamma_5 v_{\lam_2}(p_2)}
{\sqrt{2}\tilde{M_0}},
\nonumber\\
{\cal R}_{\lam_1\lam_2}^{1J_z}
&=&\frac{-\bar{u}_{\lam_1}(p_1)
\biggl[/\!\!\!\ep(J_z) -\frac{\ep\cdot(p_1-p_2)}{M_0 + m_1 + m_2}\biggr]
v_{\lam_2}(p_2)} {\sqrt{2}\tilde{M_0}},
\nonumber\\
\eea where $\tilde{M_0}=\sqrt{M^2_0-(m_1-m_2)^2}$,
$M^2_0=\sum_{i=1}^2({\bf k}^2_{i\perp}+m^2_i)/x_i$ is
the boost invariant meson mass square obtained from the free
energies of the constituents in mesons, and
$\ep^\mu(J_z)$ is the polarization vector of the vector
meson~\cite{CJ08D}. The spin-orbit wave functions satisfy the relation
$\sum_{\lam_1\lam_2}{\cal R}_{\lam_1\lam_2}^{JJ_z\dagger} {\cal
R}_{\lam_1\lam_2}^{JJ_z}=1$ for both pseudoscalar and vector
mesons. For the radial wave function $\phi$, we use the same
Gaussian wave function for both pseudoscalar and vector mesons:
\be\label{rad}
 \phi(x_i,{\bf
k}_{i\perp})=\frac{4\pi^{3/4}}{\beta^{3/2}} \sqrt{\frac{\partial
k_z}{\partial x}} {\rm exp}(-{\vec k}^2/2\beta^2),
 \ee
 where $\beta$ is the variational parameter. When the longitudinal
component $k_z$ is defined by $k_z=(x-1/2)M_0 +
(m^2_2-m^2_1)/2M_0$, the Jacobian of the variable transformation
$\{x,{\bf k}_\perp\}\to {\vec k}=({\bf k}_\perp, k_z)$ is given by
\be\label{jacob} \frac{\partial k_z}{\partial
x}=\frac{M_0}{4x_1x_2} \biggl\{ 1-
\biggl[\frac{m^2_1-m^2_2}{M^2_0}\biggr]^2\biggr\}. \ee Note that
the free kinetic part of the Hamiltonian $H_0=\sqrt{m^2_q+{\vec
k}^2}+\sqrt{m^2_{\bar{q}}+{\vec k}^2}$ is equal to the free mass
operator $M_0$ in the light-front formalism.

The normalization factor in Eq.~(\ref{rad}) is obtained from the
following normalization of the total wave function:
\be\label{norm} \int^1_0dx\int\frac{d^2{\bf k}_\perp}{16\pi^3}
|\Psi^{JJ_z}_{100}(x,{\bf k}_{i\perp})|^2=1. \ee We apply our
variational principle to the QCD-motivated effective Hamiltonian
first to evaluate the expectation value of the central Hamiltonian
$H_0+V_0$, {\em i.e.}, $\la\phi|(H_0+V_0)|\phi\ra$ with a trial
function $\phi(x_i,{\bf k}_{i\perp})$ that depends on the
variational parameter $\beta$. Once the model
parameters are fixed by minimizing the expectation value
$\la\phi|(H_0+V_0)|\phi\ra$, then the mass eigenvalue of each meson is
obtained as $M_{q\bar{q}}=\la\phi|(H_0+V_{q\bar{q}})|\phi\ra$.
Following the above procedure, we find an analytic form of the
mass eigenvalue given by
 \bea\label{Mass}
 M_{q\bar{q}}&=& \frac{{\bf 1}}{\beta\sqrt{\pi}}\sum_{i=q,\bar{q}}m^2_i
e^{m^2_i/2\beta^2}K_1(\frac{m^2_i}{2\beta^2}) + a{\bf 1} \nonumber\\
&+& b\left(
\begin{array}{c}
\frac{2}{\beta\sqrt{\pi}}\\
\frac{3}{2\beta^2}
\end{array}\,\right) -\al_s\biggl[\frac{8\beta}{3\sqrt{\pi}}
+ \frac{32\beta^3\la{\bf S}_q\cdot{\bf S}_{\bar{q}}\ra} {9m_q
m_{\bar{q}}\sqrt{\pi}}\biggr]{\bf 1}, \nonumber\\
 \eea
 where ${\bf 1}=\left(
\begin{array}{c}
1\\
1
\end{array}\,\right)$
and $K_1(x)$ is the modified Bessel function of the second
kind. The upper and lower components of the column vector in
Eq.~(\ref{Mass}) represent the results for the linear and HO
potential models, respectively. Minimizing energies with
respect to $\beta$ and searching for a fit to the observed ground
state meson spectra, our central potential $V_0$ obtained from our
optimized potential parameters ($a=-0.72$ GeV, $b=0.18$ GeV$^2$,
and $\al_s=0.31$)~\cite{CJ1} for Coulomb plus linear potential
was found to be quite comparable with the quark potential model
suggested by Scora and Isgur~\cite{SI} where they obtained
$a=-0.81$ GeV, $b=0.18$ GeV$^2$, and $\al_s=0.3\sim 0.6$ for the
Coulomb plus linear confining potential. A more detailed procedure
for determining the model parameters of light- and heavy-quark
sectors can be found in our previous works~\cite{CJ1,CJ2}. In this
work, we obtain the new variational parameter $\beta_{cb}$ for the
bottom-charm sector and predict the mass eigenvalues of the
low-lying $B_c$ and $B^*_c$ states. Our new prediction of
$M_{B_c}=6459\;[6351]$ MeV obtained from the linear [HO] potential
model is in agreement with the data, $M^{\rm exp}_{B_c}=(6276\pm
4)$ MeV~\cite{Data08} within 3$\%$ error. We also predict the
unmeasured mass of $B^*_c$ as $M_{B^*_c}=6494\;[6496]$ MeV for the
linear [HO] potential model.
Although it is generally believed that the linear potential is preferred
between quark and antiquark in the heavy meson system, our result of the
spectrum computation indicates that the HO potential is also
viable and thus leads to the further investigation.
We use both HO and linear
potentials to compute the decay processes presented below and identify
the physical observables sensitive to the choice of potential.

\begin{figure}
\vspace{1.5cm}
\includegraphics[width=3in,height=4in]{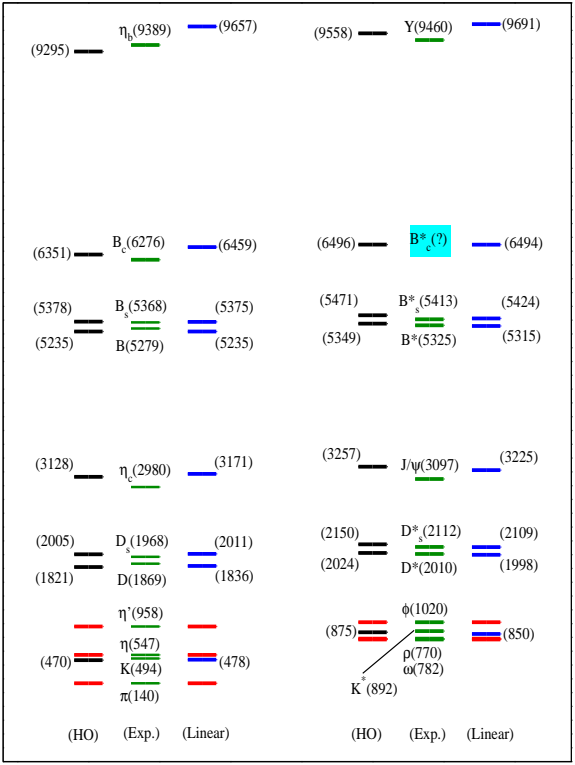}
\caption{(color online).
Fit of the ground state meson masses[MeV] with the parameters given
in Table~\ref{t1}. The $(\rho,\pi)$, $(\eta,\eta')$, and $(\omega,\phi)$
masses are our input data. The masses of $(\omega-\phi)$ and $(\eta-\eta')$
were used to determine the mixing angles of $\omega-\phi$
and $\eta-\eta'$~\cite{CJ1}, respectively.}
\label{fig1}
\end{figure}

Our model parameters $(m_q,\beta_{q\bar{q}})$ and the predictions
of the ground state meson mass spectra obtained from the linear
and HO potential models are summarized in Table~\ref{t1} and in
Fig.~\ref{fig1}, respectively, compared with the experimental
data~\cite{Data08}. Our prediction of the $\eta_b$ meson obtained from
the linear [HO] potential model, $M_{\eta_b}=9657\;[9295]$ MeV
slightly overestimates[underestimates] the very recent data from
the Babar experiment, $M^{\rm
exp}_{\eta_b}=9388.9^{+3.1}_{-2.3}(\rm stat)\pm 2.7(\rm syst)$
MeV~\cite{BaBar08}. Overall, however, our LFQM predictions of the
ground state meson mass spectra are in agreement with the
data~\cite{Data08} within 6$\%$ error.

\begin{table*}[t]
\caption{The constituent quark mass [GeV] and the Gaussian parameters
$\beta$ [GeV] for the linear and HO potentials obtained by the variational
principle. $q=u$ and $d$.}\label{t1}
\begin{tabular}{ccccccccccccccc} \hline\hline
Model & $m_q$ & $m_s$ & $m_c$ & $m_b$ &
$\beta_{qq}$ & $\beta_{qs}$  & $\beta_{ss}$ & $\beta_{qc}$
& $\beta_{sc}$ & $\beta_{cc}$ & $\beta_{qb}$ & $\beta_{sb}$
& $\beta_{cb}$ & $\beta_{bb}$ \\
\hline
Linear & 0.22 & 0.45 & 1.8 & 5.2 & 0.3659 & 0.3886 & 0.4128 &
0.4679 & 0.5016 & 0.6509 & 0.5266 & 0.5712 & 0.8068 & 1.1452\\
\hline
HO & 0.25 & 0.48 & 1.8 & 5.2 & 0.3194 & 0.3419 & 0.3681 &
0.4216 & 0.4686 & 0.6998 & 0.4960 & 0.5740 & 1.0350 &1.8025\\
\hline\hline
\end{tabular}
\end{table*}

The decay constants of pseudoscalar and vector mesons are defined
by
\bea\label{fp}
\la 0|\bar{q}\gamma^\mu\gamma_5 q|P\ra&=&if_P P^\mu,
\nonumber\\
\la 0|\bar{q}\gamma^\mu q|V(P,h)\ra&=&f_V M_V\ep^\mu(h).
\eea
In
the above definitions for the decay constants, the experimental
values of pion and rho meson decay constants are $f_\pi\approx
131$ MeV from $\pi\to\mu\nu$ and $f_\rho\approx 220$ MeV from
$\rho\to e^+e^-$.

Using the plus component ($\mu=+$) of the currents, one can easily
calculate the decay constants. The explicit forms of pseudoscalar
and vector meson decay constants are given by
\bea\label{fp_LFQM}
f_P&=&2\sqrt{6}\int \frac{dx\; d^2{\bf k}_\perp}{16\pi^3}
\frac{{\cal A}}{\sqrt{{\cal A}^2 + {\bf k}^2_\perp}}\phi(x,{\bf
k}_\perp),
\nonumber\\
f_V&=&2\sqrt{6}\int \frac{dx\; d^2{\bf k}_\perp}{16\pi^3}
\frac{\phi(x,{\bf k}_\perp)}{\sqrt{{\cal A}^2 + {\bf k}^2_\perp}}
\biggl[ {\cal A} + \frac{2{\bf k}^2_\perp}{{\cal M}_0}\biggr],
 \eea
where ${\cal A}=x_2m_1 + x_1m_2$ and ${\cal M}_0=M_0+m_1+m_2$.
Here, only $L_z=S_z=0$ component of the wave function contributes.
We note that the vector meson decay constant
$f_V$ is extracted from the longitudinal ($h=0$) polarization.

In Table~\ref{t2}, we present our predictions for the decay
constants of $f_{B_c}$ and $f_{B_c^*}$ and compare with other
model calculations~\cite{IKS01,EFG67,EQ,GKLT,Ful,CG}.
The decay constants for other light- and
heavy-mesons have been predicted in our previous
works~\cite{CJ1,Choi07,CJ08D} and found to be in good agreement with
experimental data.
While our predictions of the decay constants
for the light-light and heavy-light systems~\cite{CJ1,Choi07,CJ08D}
are not sensitive to the choice of potential(linear or HO),  the decay
constants for heavy-heavy systems such as $(B_c,B^*_c)$
and $(\eta_b,\Upsilon)$ in~\cite{Choi07}
are quite sensitive to the choice of potential. Thus, the experimental
measurements for the decay constants of $(B_c, B^*_c)$ and $(\eta_b,\Upsilon)$
mesons may distinguish between the linear and HO potentials within our LFQM.
\begin{table}[t]
\caption{Bottom-charm meson decay constants(in unit of MeV)
obtained from the linear [HO] parameters.}\label{t2}
\begin{tabular}{cccccccc} \hline\hline
 & Linear[HO] &~\cite{IKS01} &~\cite{EFG67} & ~\cite{EQ} & ~\cite{GKLT} & ~\cite{Ful} & ~\cite{CG}
\\
\hline $f_{B_c}$  & 377[508] & 360 & 433 & 500 & $460\pm 60$ & 517
&
$410\pm 40$  \\
\hline $f_{B^*_c}$  & 398[551] &$-$& 503 & 500 & $460\pm 60$ & 517 &$-$ \\
\hline\hline
\end{tabular}
\end{table}

\begin{figure}
\vspace{1cm}
\includegraphics[width=3.0in,height=2.5in]{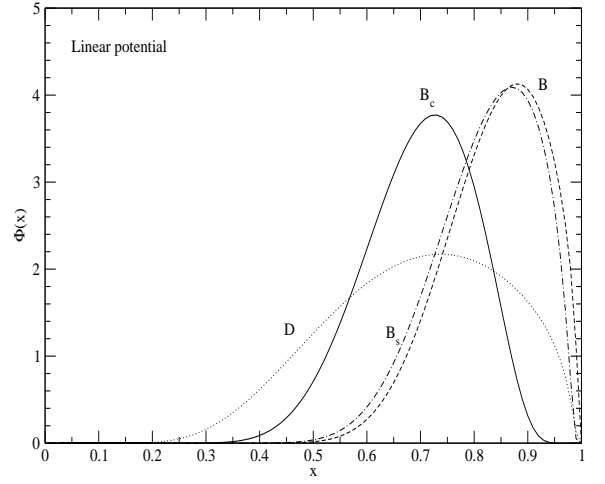}
\\\vspace{1.2cm}
\includegraphics[width=3.0in,height=2.5in]{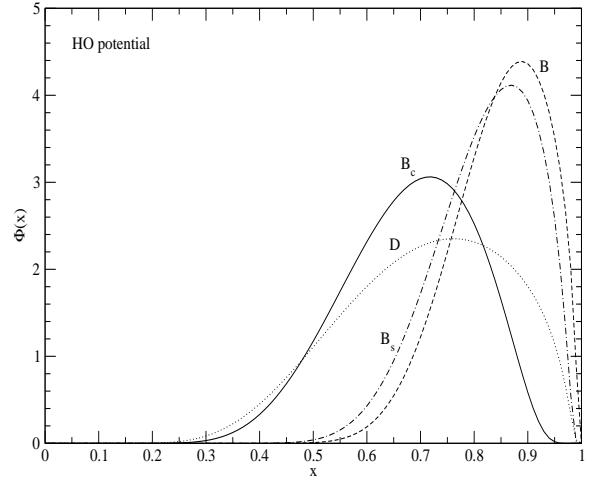}\\
\caption{The normalized distribution amplitudes for $D$, $B$,
$B_s$, and $B_c$ mesons obtained from the linear (upper panel) and
HO (lower panel) potential parameters.} \label{fig2}
\end{figure}

\begin{figure}
\vspace{1cm}
\includegraphics[width=3.0in,height=2.5in]{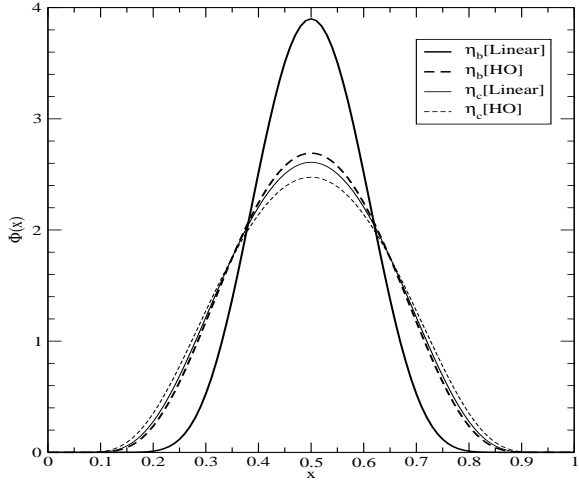}
\caption{The normalized distribution amplitudes for $\eta_c$ (thin
lines) and $\eta_b$ (thick lines) mesons obtained from the
linear (solid lines) and HO (dashed lines) potential parameters.}
\label{fig3}
\end{figure}

The process-independent quark distribution amplitude(DA) $\phi_{P(V)}(x)$ for pseudoscalar
(vector) meson is the probability amplitude for finding the $q\bar{q}$ pair in the meson
with $x_q=x$ and $x_{\bar{q}}=1-x$. It is directly related to our LF valence wave
function~\cite{CJ08D}:
\be\label{DA0}
\phi_{P(V)}= \int \frac{d^2{\bf k}_\perp}{16\pi^3}
\Psi(x,{\bf k}_\perp).
\ee
The ${\bf k}_\perp$ integration in Eq.~(\ref{DA0}) is cut off by the ultraviolet cutoff
$\Lambda$ implicit in the wave function. The dependence on the scale $\Lambda$ is then given
by the QCD evolution equation~\cite{LB} and can be calculated perturbatively.
However, the DAs at a certain low scale can be obtained by the necessary nonperturbative input
from LFQM. Moreover, the presence of the damping Gaussian factor in our LFQM allows us to perform
the integral up to infinity without loss of accuracy. The quark DAs for pseudoscalar and vector
mesons are constrained by
 \be\label{DA}
\int^1_0\phi_{P(V)}(x)dx=\frac{f_{P(V)}}{2\sqrt{6}}.
 \ee
%
 We show in Fig.~\ref{fig2} the normalized quark DAs
$\Phi(x)=(2\sqrt{6}/f_P)\phi(x)$ for $D$ (dotted line), $B$ (dashed
line), $B_s$ (dot-dashed line), and $B_c$ (solid line) mesons
obtained from the linear (upper panel) and HO (lower panel) potential parameters,
respectively. In Fig.~\ref{fig3}, we also show the normalized quark
DAs for $\eta_c$ (thin lines) and $\eta_b$ (thick lines) mesons
obtained from the linear(solid lines) and HO(dashed lines)
potential parameters. While the two model
predictions for the heavy-light systems such as ($D$, $B$, and
$B_s$) are not much different from each other, the HO potential model
predictions for the heavy-heavy systems such as ($\eta_c$, $B_c$
and $\eta_b$) give somewhat broader shapes than the linear
potential model predictions.

\section{Semileptonic decays of the $B_c$ meson}
The relevant quark momentum variables for $P(q_1\bar{q})\to P(q_2\bar{q'})$
transitions in the $q^+=0$ frames are given by
 \bea\label{mom}
p^+_1&=& x_1P^+_1, \;\; p^+_{\bar{q}}=x_2P^+_1, \;\;
{\bf p}_{1\perp}=x_1{\bf P}_{1\perp}-{\bf k}_\perp,
\nonumber\\
p^+_2&=& x_1P^+_1, \;\; p^+_{\bar{q'}}=x_2P^+_1, \;\;
{\bf p}_{2\perp}=x_1{\bf P}_{2\perp}-{\bf k}'_\perp,
\nonumber\\
{\bf p}_{{\bar q}\perp}&=& x_2{\bf P}_{1\perp}+{\bf k}_\perp,\;\;
{\bf p}_{{\bar q'}\perp}=x_2{\bf P}_{2\perp}+{\bf k}'_\perp,
 \eea
 where $x_1=x$ and $x_2=1-x$ and spectator quark requires that
 $p^+_{\bar{q}}=p^+_{\bar{q'}}$ and
 ${\bf p}_{{\bar q}\perp}={\bf p}_{{\bar q'}\perp}$.
 Taking a Lorentz frame where ${\bf P}_{1\perp}=0$ and
 ${\bf P}_{2\perp}=-{\bf q}_\perp$ amounts to
 ${\bf k'}_\perp={\bf k}_\perp + (1-x){\bf q}_\perp$.

The hadronic matrix element of the plus current
${\cal M}^+\equiv\la P_2|V^+|P_1\ra$
in Eq.~(\ref{PPF}) is then obtained by the convolution formula of the
initial and final state LF wave functions in the valence region:
 \bea\label{fp0}
 {\cal M}^+ &=&
 \int^{1}_{0}dx\int
\frac{d^{2}{\bf k}_{\perp}}{16\pi^3}
\phi_{2}(x,{\bf k'}_{\perp})\phi_{1}(x,{\bf k}_{\perp})\nonumber\\
&&\times\;\sum_{\lambda_1,\lambda_2,\bar{\lambda}}
{\cal R}^{00\dagger}_{\lambda_2\bar{\lambda}}
\frac{\bar{u}_{\lambda_2}(p_2)}{\sqrt{p^+_2}}\gamma^+
\frac{u_{\lambda_1}(p_1)}{\sqrt{p^+_1}}
{\cal R}^{00}_{\lambda_1\bar{\lambda}}.
 \eea
Substituting the covariant form of the spin-orbit wave function for
pseudoscalar meson given by Eq.~(\ref{R00_A}) into Eq.~(\ref{fp0})
yields
 \bea\label{fp2}
 \hspace{-0.5cm} {\cal M}^+ &=& -\int^{1}_{0}dx\int
\frac{d^{2}{\bf k}_{\perp}}{16\pi^3}
\frac{\phi_{2}(x,{\bf k'}_{\perp})\phi_{1}(x,{\bf k}_{\perp})}
{2xP^+_1\tilde{M}_0\tilde{M'}_0}\nonumber\\
&\times&\hspace{-0.0cm}
{\rm Tr}[\gamma^5(\not\!p_2+m_2)\gamma^+(\not\!p_1
+m_1)\gamma^5(\not\!p_{\bar{q}}-m_{\bar{q}})],
\eea
where
 \bea\label{M00}
 \tilde{M}_0&=& \sqrt{M^2_0 -(m_1-m^2_{\bar{q}})^2}
 =\sqrt{\frac{{\bf k}^2_\perp + {\cal A}^2_1}{x(1-x)}},
 \nonumber\\
 \tilde{M'}_0&=& \sqrt{M'^2_0 -(m_2-m^2_{\bar{q}})^2}
 =\sqrt{\frac{{\bf k'}^2_\perp + {\cal A}^2_2}{x(1-x)}},
 \eea
 and  ${\cal A}_{i}= (1-x) m_{i} + x m_{\bar{q}}$ ($i=1,2$).
 After some manipulation,
 the trace term in Eq.~(\ref{fp2}) is reduced to
 \bea\label{tr0}
 \nonumber\\
 &&{\rm Tr}[\gamma^5(\not\!p_2+m_2)\gamma^+(\not\!p_1
 +m_1)\gamma^5(\not\!p_{\bar q}-m)]
 \nonumber\\
 &&= -\frac{4P^+_1}{(1-x)}({\bf k}_\perp\cdot{\bf k}'_\perp
 + {\cal A}_1 {\cal A}_2 ).
 \eea
Finally, the form factor $f_+(q^2)$ obtained from the valence contribution in
the $q^+=0$ frame is given by
 \bea\label{fpp}
 f_{+}(q^2) &=& \int^{1}_{0}dx\int
\frac{d^{2}{\bf k}_{\perp}}{16\pi^3}
\frac{\phi_{1}(x,{\bf k}_{\perp})}{\sqrt{ {\cal A}_{1}^{2}
 + {\bf k}^{2}_{\perp}}}
\frac{\phi_{2}(x,{\bf k}'_{\perp})}{\sqrt{ {\cal A}_{2}^{2}
+ {\bf k}^{\prime 2}_{\perp}}}
\nonumber\\
 && \hspace{1cm} \times\;
 ( {\cal A}_{1}{\cal A}_{2}+{\bf k}_{\perp}\cdot{\bf k'}_{\perp} ).
 \eea
We should note in the trace calculation of Eq.~(\ref{tr0}) that the internal momenta of the
valence quarks carried inside mesons are all on-mass-shell ($p^2_i=m^2_i$).
Nevertheless, the LF valence contribution to the form factor $f_+(q^2)$ is shown to
be equivalent to the covariant result as shown in Sec.~II.

Comparing the manifestly covariant form factor $f_+(q^2)$
in Eq.~(\ref{ap:23}) and our LFQM result $f_+(q^2)$ in Eq.~(\ref{fpp}),
we find the following relations for the LF vertex functions between the two models:
\bea\label{ap:33}
 \sqrt{2N} \frac{ \chi_1(x,{\bf k}_\perp) } {1-x}
 &=& \frac{ \phi_1(x,{\bf k}_\perp) } {\sqrt{ {\cal A}^2_1 + {\bf k}^2_\perp }},
\nonumber\\
\sqrt{2N} \frac{ \chi_2(x,{\bf k'}_\perp) } {1-x}
 &=& \frac{ \phi_2(x,{\bf k'}_\perp) } {\sqrt{ {\cal A}^2_2 + {\bf k}^{\prime 2}_\perp }}.
 \eea
We should note that the zero-mode operator included in Eq.~(\ref{ap:29}) is independent
from the choice of radial wave function.

Applying the relation in Eq.~(\ref{ap:33}) to  Eqs.~(\ref{ap:23}) and~(\ref{ap:29}),
we get the following LF valence contribution $f^{val}_-(q^2)$ and the LF covariant
solution $f^{full}_-(q^2)$ including both the valence and the zero-mode contributions within
our LFQM:

 \begin{widetext}
 \bea\label{fmv_LFQM}
  f^{val}_-(q^2) &=& \int^1_0 (1-x) dx
  \int \frac{ d^2{\bf k}_\perp } { 16\pi^3 }
  \frac{ \phi_1 (x, {\bf k}_\perp) } {\sqrt{ {\cal A}^2_1 + {\bf k}^2_\perp }}
  \frac{ \phi_2 (x, {\bf k'}_\perp) } {\sqrt{ {\cal A}^2_2 + {\bf k}^{\prime 2}_\perp }}
  \biggl\{ - (1-x) M^2_1 + (m_2 - m_{\bar{q}}){\cal A}_1 - m_{\bar{q}}(m_1 - m_{\bar{q}})
  \nonumber\\
  && + \frac{ {\bf k}_\perp\cdot{\bf q}_\perp }{q^2} [ M^2_1 + M^2_2
  - 2(m_1 - m_{\bar{q}}) (m_2 - m_{\bar{q}}) ]
  \biggr\},
  \nonumber\\
    \nonumber\\
 f^{full}_-(q^2) &=& \int^1_0 (1-x) dx
 \int \frac{ d^2{\bf k}_\perp } { 16\pi^3 }
  \frac{ \phi_1 (x, {\bf k}_\perp) } {\sqrt{ {\cal A}^2_1 + {\bf k}^2_\perp }}
  \frac{ \phi_2 (x, {\bf k'}_\perp) } {\sqrt{ {\cal A}^2_2 + {\bf k}^{\prime 2}_\perp }}
   \biggl\{ -x(1-x) M^2_1 - {\bf k}^2_\perp - m_1m_{\bar{q}} + (m_2 - m_{\bar{q}}){\cal A}_1
 \nonumber\\
 && + 2\frac{q\cdot P}{q^2} \biggl[ {\bf k}^2_\perp
 + 2\frac{ ( {\bf k}_\perp \cdot {\bf q}_\perp)^2 } {q^2} \biggr]
 + 2 \frac{ ( {\bf k}_\perp \cdot {\bf q}_\perp)^2 } {q^2}
  + \frac{ {\bf k}_\perp \cdot {\bf q}_\perp } {q^2}  [ M^2_2 - (1-x) (q^2 + q\cdot P) + 2 x M^2_0
  \nonumber\\
  && - (1 - 2x) M^2_1 - 2(m_1 - m_{\bar{q}}) (m_1 + m_2) ] \biggr\},
  \eea
  \end{widetext}
where $M_1$ and $M_2$ are the physical masses of the initial and final mesons, respectively.

\section{Radiative $B^*_c\to B_c\gamma$ decay}

In addition to semileptonic decays, the radiative decays of vector
mesons can be anlayzed within our LFQM~\cite{CJ1,Choi07}.
In this work, we thus calculate the decay
rate
for
$B^*_c\to B_c\gamma$ transition.

\begin{figure}
\vspace{0.8cm}
\includegraphics[width=3in,height=1.5in]{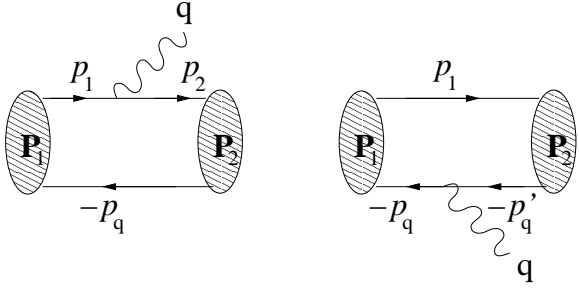}
\caption{The lowest order Feynman diagram for $V\to P\gamma^*$ process.}
\label{fig5}
\end{figure}

In our LFQM calculation of $B^*_c\to B_c\gamma$ process, we first
analyze the virtual photon ($\gamma^*$) decay process,
calculating the momentum dependent transition form factor,
$F_{B^*_cB_c}(q^2)$. The lowest-order Feynman diagram for
$V\to P\gamma^*$ process is shown in Fig.~\ref{fig5} where the decay from
vector meson to pseudoscalar meson and virtual photon state
is mediated by a quark loop with flavors of constituent mass
$m_1$ and $m_{\bar q}$.

The transition form factor
$F_{B^*_cB_c}(q^2)$ for $V(P_1)\to P(P_2)+\gamma^*(q)$ is
defined as~\cite{Choi07}
\be\label{ff}
\la P(P_2)|V^\mu|V(P_1,h)\ra
=ie\epsilon^{\mu\nu\rho\sigma}\epsilon_\nu(P_1,h)
q_{\rho}P_{1\sigma}F_{VP}(q^2),
 \ee
 where the antisymmetric
tensor $\ep^{\mu\nu\rho\sigma}$ assures electromagnetic gauge
invariance, $q=P_1-P_2$ is the four-momentum of the virtual
photon, and
$\ep_\nu(P_1,h)$ is the polarization vector of the initial
meson with the four-momentum $P_1$ and the helicity $h$. The kinematically
allowed $q^2$ (momentum transfer squared) ranges from 0 to $q^2_{\rm
max}=(M_{V}-M_{P})^2$. The decay form factor
$F_{B^*_cB_c}(q^2)$ can also be obtained in the $q^+=0$ frame with
the transverse ($h=\pm 1$) polarization and the ``$+$"-component of
the currents
without encountering zero-mode contributions~\cite{Zero}
and then analytically continued from
the spacelike region where the form factor is given by
$F_{B^*_cB_c}({\bf q}^2_\perp)$ to the timelike $q^2>0$
region by changing ${\bf q}^2_\perp$ to $-q^2$ in the form factor.

The hadronic matrix element of the plus current
${\cal M}^+_{VP}\equiv\la P(P_2)|V^+|V(P_1,h)\ra$
in Eq.~(\ref{ff}) is then obtained by the convolution formula of the
initial and final state LF wave functions in the valence region:
 \bea\label{ff2}
 {\cal M}^+_{VP} &=& \sum_j ee_j
 \int^{1}_{0}dx\int
\frac{d^{2}{\bf k}_{\perp}}{16\pi^3}
\phi_{2}(x,{\bf k'}_{\perp})\phi_{1}(x,{\bf k}_{\perp})\nonumber\\
&&\times\;\sum_{\lambda_1,\lambda_2,\bar{\lambda}}
{\cal R}^{00\dagger}_{\lambda_2\bar{\lambda}}
\frac{\bar{u}_{\lambda_2}(p_2)}{\sqrt{p^+_2}}\gamma^+
\frac{u_{\lambda_1}(p_1)}{\sqrt{p^+_1}}
{\cal R}^{11}_{\lambda_1\bar{\lambda}},
 \eea
 where $ee_j$ is the electrical charge for $j$th quark flavor.
Substituting the covariant forms of the spin-orbit wave
functions
for pseudoscalar and vector mesons given by Eq.~(\ref{R00_A})
into Eq.~(\ref{ff2}) and comparing
it
with the right-hand side of
Eq.~(\ref{ff}), i.e. $eP^+_1F_{VP}(q^2)q^R/\sqrt{2}$ where
$q^R=q_x + iq_y$, we could extract the one-loop integral,
$I(m_1,m_{\bar q},q^2)$,
given by
\bea\label{soft_form}
I(m_1,m_{\bar q},q^2) \hspace{-0.2cm} &=&\hspace{-0.2cm}
\int^1_0 dx\int \frac{d^2{\bf k}_\perp}{8\pi^3}
 \frac{ \phi_1 (x, {\bf k}_\perp) } {\sqrt{ {\cal A}^2_1 + {\bf k}^2_\perp }}
  \frac{ \phi_2 (x, {\bf k'}_\perp) } {\sqrt{ {\cal A}^2_1 + {\bf k}^{\prime 2}_\perp }}
\nonumber\\
&\times& (1-x) \biggl\{{\cal A}_1 + \frac{2} {{\cal M}_0} [{\bf k}^2_\perp
+ \frac{({\bf k}_\perp\cdot{\bf q}_\perp)^2}{q^2}]
\biggr\},
\nonumber\\
\eea
where ${\cal M}_0=M_0 + m_1 + m_{\bar q}$.
The decay form factor $F_{B^*_cB_c}(q^2)$ is then obtained
as~\cite{Choi07}
 \be\label{FS} F_{B^*_cB_c}(q^2)=
e_1I(m_1,m_{\bar q},q^2) + e_2 I(m_{\bar q},m_1,q^2).
\ee
The coupling constant $g_{B^*_cB_c\gamma}$ for real
photon ($\gamma$) case can then be determined in the limit
$q^2\to 0$, i.e., $g_{B^*_cB_c\gamma}=F_{B^*_cB_c}(q^2=0)$. The
decay width for $V\to P\gamma$ is given by \be\label{width}
\Gamma(B^*_c\to B_c\gamma)=\frac{\alpha}{3}g_{B^*_cB_c\gamma}^2
k^3_\gamma, \ee where $\alpha$ is the fine-structure constant  and
$k_\gamma=(M^2_{B^*_c}-M^2_{B_c})/2M_{B^*_c}$ is the kinematically
allowed energy of the outgoing photon.

\section{Numerical Results}

\begin{figure}
\vspace{0.8cm}
\includegraphics[width=3in,height=2.2in]{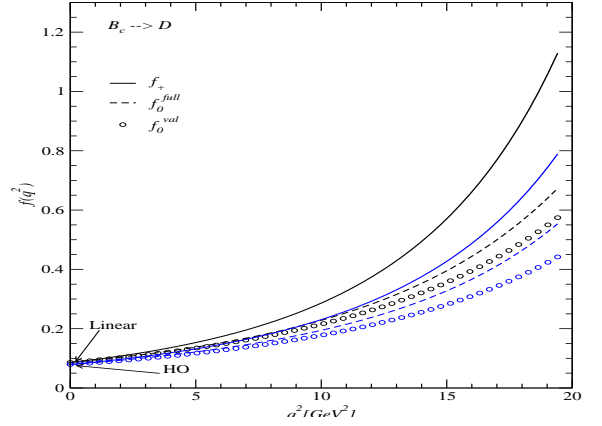}\\
\vspace{1cm}
\includegraphics[width=3in,height=2.2in]{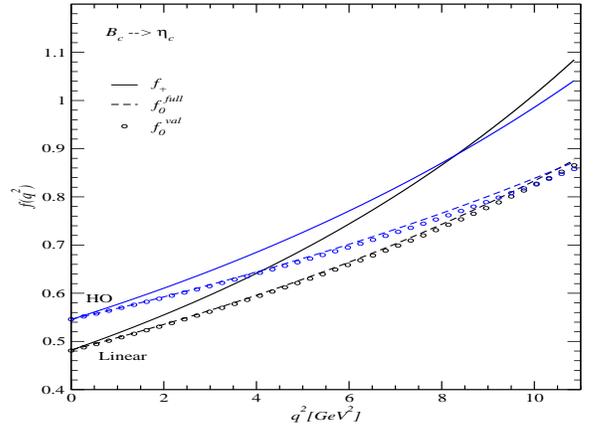}
\caption{(color online). The weak form factors $f_+(q^2)$ (solid
line) and $f_0(q^2)$ (dashed line) for $B_c\to D$ (upper panel) and
$B_c\to\eta_c$ (lower panel) semileptonic decays obtained from the
linear (black line) and HO (blue line) potential parameters. The circles
represent the valence contributions $f^{val}_0(q^2)$ to $f_0(q^2)$.}
\label{fig6}
\end{figure}

\begin{figure}
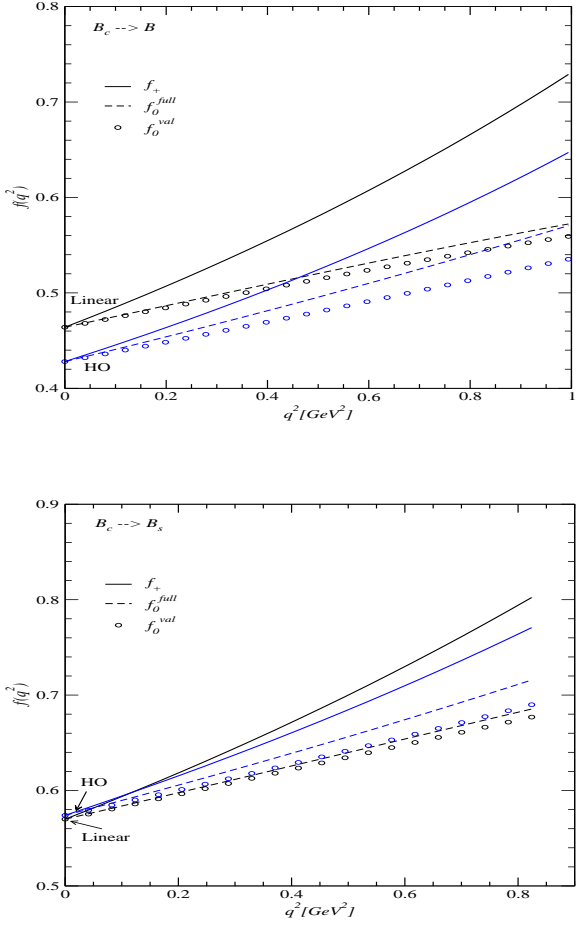

\vspace{0.8cm}
\includegraphics[width=3in,height=2.2in]{fig7a.eps}\\
\vspace{1cm}
\includegraphics[width=3in,height=2.2in]{fig7b.eps}
\caption{(color online). The weak form factors $f_+(q^2)$ and
$f_0(q^2)$ for $B_c\to B$ (upper panel) and $B_c\to B_s$ (lower
panel) semileptonic decays obtained from the linear and HO
potential parameters. The same line codes are used as in
Fig.~\ref{fig6}.} \label{fig7}
\end{figure}

\begin{figure}
\vspace{0.8cm}
\includegraphics[width=3in,height=2.2in]{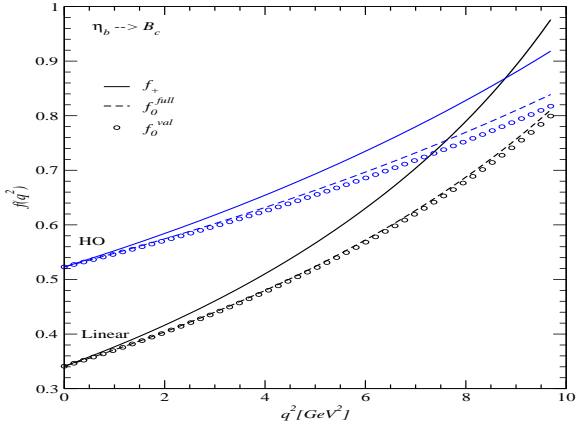}
\caption{(color online). The weak form factors $f_+(q^2)$ and
$f_0(q^2)$ for $\eta_b\to B_c$ semileptonic decay obtained from
the linear and HO potential parameters. The same line codes are
used as in Fig.~\ref{fig6}.} \label{fig8}
\end{figure}

In our numerical calculations of exclusive $B_c$ decays, we use
two sets of model parameters ($m,\beta$) for the linear and HO
confining potentials given in Table~\ref{t1} obtained
from the calculation of the mass spectra.
Although our predictions of ground state heavy
meson masses are overall in good agreement with the experimental
values, we use the experimental meson masses~\cite{Data08} in the
computations of the decay widths to reduce possible theoretical
uncertainties. We also use the central values of the CKM matrix elements,
\bea\label{CKMEQ} |V_{ub}|&=& 0.00393, \;\;\hspace{1cm}
|V_{cb}|=0.0412,
\nonumber\\
|V_{cd}|&=& 0.230, \;\;\;\; \;\hspace{1.1cm} |V_{cs}|=1.04, \eea
quoted by the Particle Data Group (PDG)~\cite{Data08}.

In Figs.~\ref{fig6} and~\ref{fig7} we show the $q^2$-dependence
of the
LF covariant
weak form factors $f_+(q^2)$(solid lines) and $f_0(q^2)$(dashed lines)
in the whole kinematical
ranges
for the CKM-suppressed (enhanced) semileptonic
$B_c\to D(\eta_c)$ (Fig.~\ref{fig6}) and $B_c\to
B(B_s)$ (Fig.~\ref{fig7}) decays obtained from both linear(black lines)
and HO (blue lines) potential models. The
circles
represent the valence contributions $f^{val}_0(q^2)$ to $f_0(q^2)$.
That is, the difference between $f_0(q^2)$ and
$f^{val}_0(q^2)$ represents the zero-mode contribution to $f_0(q^2)$.

The kinematical ranges for $B_c\to D(\eta_c)$ decays induced by
$b\to u(c)$ transitions with the $c$ quark being a spectator are
considerably broader than those for $B_c\to B(B_s)$ decays induced
by $c\to d(s)$ transitions with the $b$ quark being a spectator.
The form factors $f_+(q^2)$ and $f_0(q^2)$ at the zero-recoil
point (i.e., $q^2=q^2_{\rm max})$ correspond to the overlap integral
of the initial and final state meson wave functions. The
maximum-recoil point (i.e., $q^2=0$) corresponds to a final state
meson recoiling with the maximum three-momentum $|{\vec
P}_f|=(M^2_{B_c}-M^2_f)/2M_{B_c}$ in the rest frame of the $B_c$
meson. Especially for the $B_c\to D$ decay, the light $\bar{u}$
quark in $D$ meson will typically recoil with the momentum comparable
to or larger than the $c$ quark mass due to the large recoil
effect for $B_c\to D$ decay. In order for the final $D$ meson to
be bound, there must be a correspondingly large momentum transfer
to the spectator $c$ quark. Thus, the overlap between the initial
and final meson wave functions at the maximum-recoil point is
limited and yields smaller value of $f_+(0)$ for $B_c\to
D$ decay than that for other processes. We also note that one
cannot apply the heavy quark symmetry to the system with the
two heavy quarks, due to the flavor symmetry breaking by
the kinetic energy terms as discussed in~\cite{JLMS}.
As for the zero-mode contributions, we find that the zero-mode
contributions to
$f_0(q^2)$(or $f_-(q^2)$)
 for the $B_c\to D, B_c\to B$
and $B_c\to B_s$ processes are relatively larger than that for the
$B_c\to \eta_c$ process.
For the $B_c\to \eta_c$ transition, the zero-mode contributions
to $f_0(q^2)$ obtained from both linear and HO potential
models are almost suppressed in the whole kinematical range
and moreover
the values of $f_0(q^2_{\rm max})$
almost converge to a single value regardless of the choice of potential.

Although there already exist various model predictions on the
above $B_c$ semileptonic decays, the predictions of the
semileptonic $\eta_b\to B_c$ decay is not reported yet as far as
we know. We thus show in Fig.~\ref{fig8} the $q^2$-dependence of
the weak form factors $f_+(q^2)$ and $f_0(q^2)$ for the
semileptonic $\eta_b\to B_c$ decay obtained from
both linear and HO potential models.
The same line codes presented in Fig.~\ref{fig6}
are used in Fig.~\ref{fig8}. While the linear and HO potential models give
similar decay constants for heavy-light mesons ($D,B,B_s$) and
$\eta_c$ meson~\cite{Choi07}, they predict quite different values
of $B_c$ and $\eta_b$, e.g., $f_{\eta_b}=507$ MeV and 897 MeV for
the linear and HO potential models~\cite{Choi07}, respectively.
This results in sizable differences between
the two models
for the predictions of
$f_+(q^2)$ and $f_0(q^2)$ in the $\eta_b\to B_c$ decay.
Since the linear
potential model prediction of the
quark DA for $\eta_b$ is narrower than the HO model
prediction (see Fig.~\ref{fig3}),
the overlap between the initial and final meson wave functions at
the maximum-recoil point (i.e., $q^2=0$) produces smaller values of
$f_+(=f_0)$ for the linear potential model than for the HO
potential model.
The experimental measurement of this process may also distinguish
between the linear and HO potential models within our LFQM.
The zero-mode contribution to
$f_-(q^2)$(or $f_0(q^2)$) is again quite suppressed
in the whole
kinematical range
as in the case of $B_c\to\eta_c$ process.

\begin{figure}
\vspace{1cm}
\includegraphics[width=3in,height=2.2in]{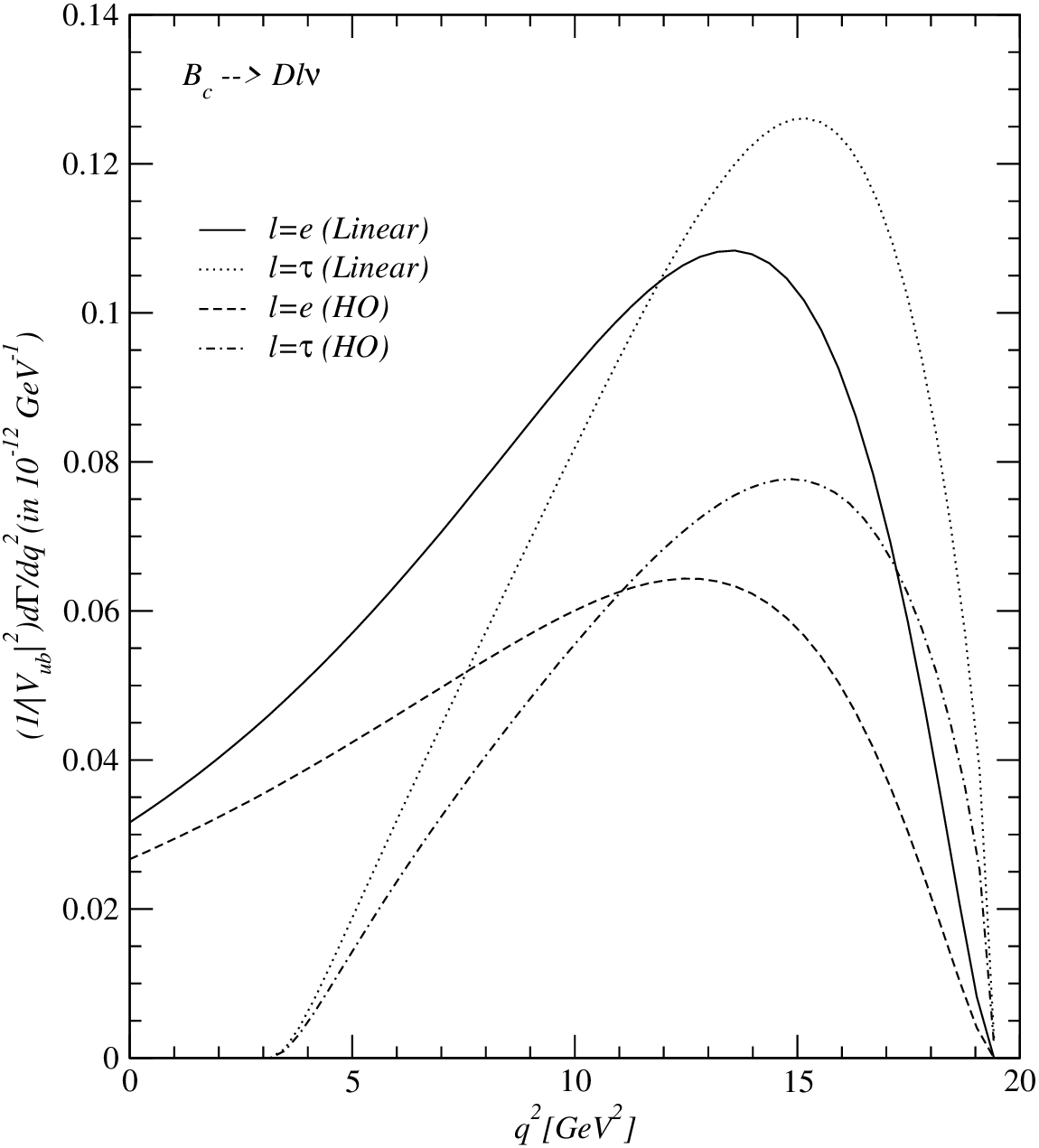}\\
\vspace{1cm}
\includegraphics[width=3in,height=2.2in]{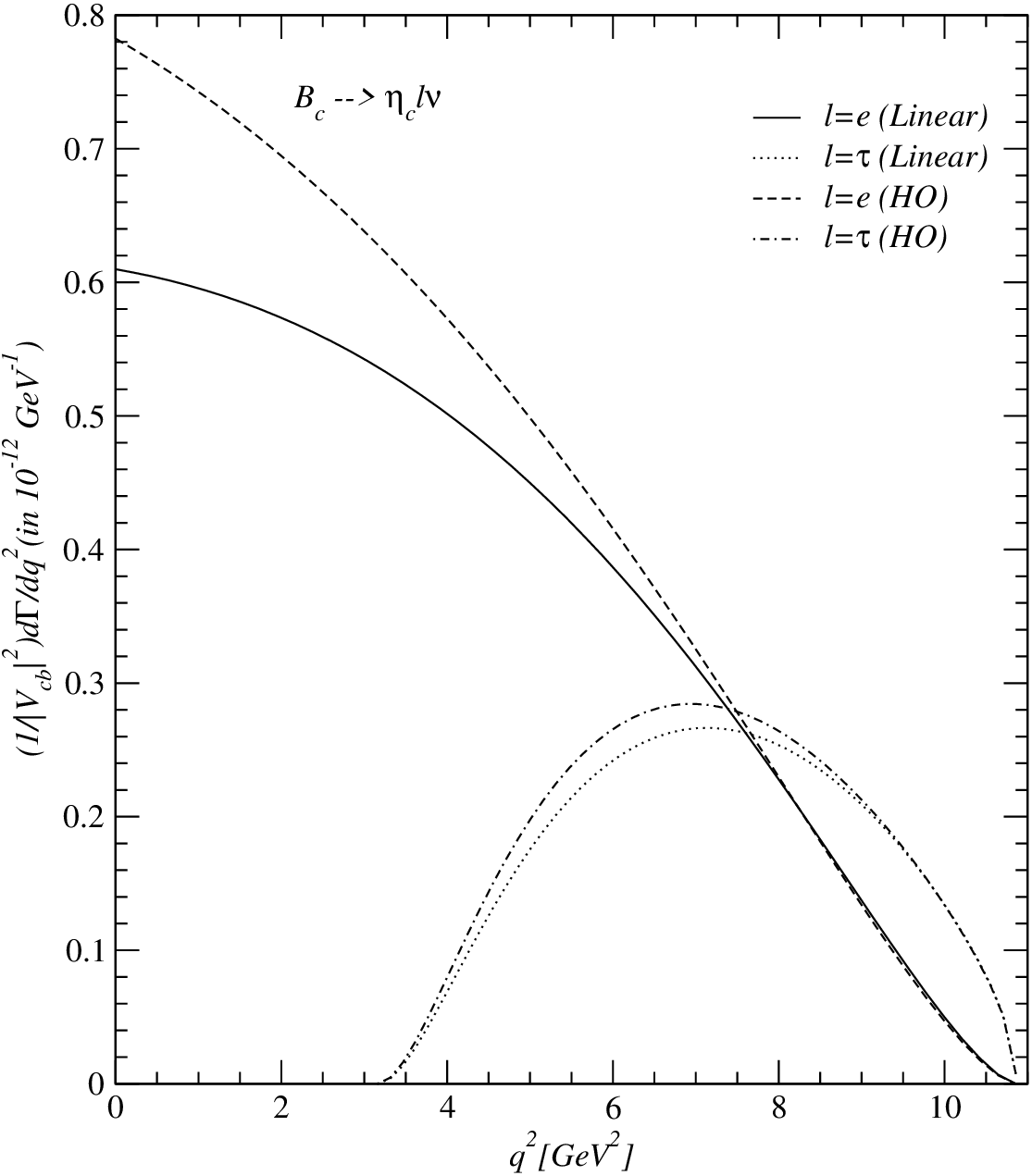}\\
\caption{Differential decay widths
$(1/|V_{Q_1Q_2}|^2)d\Gamma/dq^2)$(in units of $10^{-12}$
GeV$^{-1}$) for $B_c\to D \ell\nu_\ell$ and $B_c\to\eta_c
\ell\nu_\ell$ processes obtained from the linear and HO potential
parameters.} \label{fig9}
\end{figure}

\begin{figure}
\vspace{1cm}
\includegraphics[width=3in,height=2.2in]{fig10a.eps}\\
\vspace{1cm}
\includegraphics[width=3in,height=2.2in]{fig10b.eps}
\caption{Differential decay widths
$(1/|V_{Q_1Q_2}|^2)d\Gamma/dq^2)$(in units of $10^{-12}$
GeV$^{-1}$) for $B_c\to B \ell\nu_\ell$ and $B_c\to B_s
\ell\nu_\ell$ processes obtained from the linear and HO potential
parameters.} \label{fig10}
\end{figure}

\begin{figure}
\vspace{1cm}
\includegraphics[width=3in,height=2.2in]{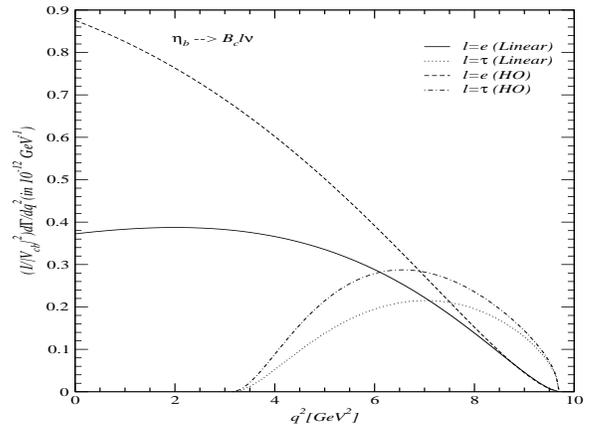}
\caption{Differential decay width
$(1/|V_{Q_1Q_2}|^2)d\Gamma/dq^2)$(in units of $10^{-12}$
GeV$^{-1}$) for  $\eta_b\to B_c \ell\nu_\ell$ process obtained
from the linear and HO potential parameters.} \label{fig11}
\end{figure}

In Figs.~\ref{fig9}-\ref{fig11}, we show the differential decay
widths $d\Gamma/dq^2$ for the  $B_c\to D(\eta_c)
\ell\nu_\ell$ (Fig.~\ref{fig9}), $B_c\to B(B_s)
\ell\nu_\ell$ (Fig.~\ref{fig10}) and $\eta_b\to B_c
\ell\nu_\ell$ (Fig.~\ref{fig11}) processes obtained from the linear
and HO potential parameters.  The line codes are described in each
figure. We should note that the minimum $q^2$ value of the form
factor depends on the actual final lepton and it is given
(neglecting neutrino masses) by the lepton mass as $q^2_{\rm
min}=m^2_{\ell}$. Although the difference between the linear and
HO model predictions are not very large for the $B_c\to (B,B_s)$
processes, they are quite different for other processes,
especially for the $\eta_b\to B_c$ process. Since the
constituent masses of $b$- and $c$ quarks are common to
both linear and HO potential models,
the difference of the decay rates for the $\eta_b\to
B_c$ process seems to come from the different choice of the variational
$\beta$ parameters. We note, however, that the difference of the decay
rates between the two models are significantly reduced for the
heavy $\tau$ lepton case.

\begin{table*}[t]
\caption{Form factors $f_+$ and $f_0$ evaluated at $q^2=0$ and
$q^2_{\rm max}$ and decay widths $\Gamma_{\ell}$(in $10^{-15}$
GeV) for $B_c\to (D,\eta_c, B, B_s)\ell\nu_{\ell}$ and $\eta_b\to
B_c \ell\nu_{\ell}(\ell=e,\mu,\tau)$ transitions.}\label{t3}
\begin{tabular}{cccccccccc} \hline\hline
Mode &  &  Linear[HO] & EFG~\cite{EFG03D,EFG03E} &
IKS~\cite{IKS01} &
NW~\cite{NW}& HNV~\cite{HNV} & AKN~\cite{AKN} & CD~\cite{CD00} & WSL~\cite{WSL} \\
\hline $B_c\to D$ & $f_{+(0)}(0)$ & 0.086[0.079] & 0.14 & 0.69& 0.1446 & - & 0.089 & - & 0.16 \\
& $f_+(q^2_{\rm max})$ & 1.129[0.789]    & 1.20& 2.20 & 1.017 & - & - & 0.59 & 1.10 \\
& $f_0(q^2_{\rm max})$ & 0.673[0.554]    & 0.64& - & - & - & - & - & 0.59 \\
& $\Gamma_{e(\mu)}$             & 0.021[0.014] & 0.019 & 0.26 & 0.020 & - & - &0.005(0.03)& 0.043\\
& $\Gamma_\tau$             & 0.019[0.012] & - & - & - & - & - & - & -\\
\hline $B_c\to\eta_c$ & $f_{+(0)}(0)$ & 0.482[0.546] & 0.47 &
0.76& 0.5359 & 0.49 & 0.622 & - & 0.61 \\
& $f_+(q^2_{\rm max})$ & 1.084[1.035] & 1.07
& 1.07 & 1.034 & 1.00 & - & 0.94 & 1.10 \\
& $f_0(q^2_{\rm max})$ & 0.876[0.872] & 0.92
& - & - & 0.91 & - & - & 0.86\\
& $\Gamma_{e(\mu)}$ & $ 6.93[7.95]$ & 5.9 & 14.0 & 6.8 & 6.95 &
8.6 &
2.1(6.9) & 9.81\\
& $\Gamma_\tau$ & 2.31[2.46] & - & 3.52 & - &
2.46 & $3.3\pm0.9$ &
- & -\\
 \hline $B_c\to B$ & $f_{+(0)}(0)$ & 0.464[0.428] & 0.39 &
0.58& 0.4504 & 0.39 & 0.362 & - & 0.63 \\
& $f_+(q^2_{\rm max})$ & 0.729[0.647] & 0.96
& 0.96 & 0.6816 & 0.70 & - & 0.66 & 0.97 \\
& $f_0(q^2_{\rm max})$ & 0.572[0.570] & 0.80
& - & - & 0.71 & - & - & 0.81 \\
& $\Gamma_{e}$
& $ 0.84[0.69]$ & 0.6 & 2.1 & 0.638 & 0.65 & - & 0.9(1.0) & 1.63\\
& $\Gamma_{\mu}$
& 0.80[0.67] & - & - & - & 0.63 & - & - & -\\
 \hline $B_c\to B_s$ & $f_{+(0)}(0)$ & 0.570[0.574] & 0.50 &
0.61& 0.5917 & 0.58 & 0.564 & - & 0.73 \\
& $f_+(q^2_{\rm max})$ & 0.802[0.771] & 0.99
& 0.92 & 0.8075 & 0.86 & - & 0.66 & 1.03 \\
& $f_0(q^2_{\rm max})$ & 0.685[0.716] & 0.86
& - & - & 0.86 & - & - & 0.87 \\
& $\Gamma_e$
& $ 15.45[15.20]$ & 12 & 29 & 12.35 & 15.1 & 15 & 11.1(12.9) & 23.45\\
& $\Gamma_\mu$
& $ 14.61[14.40]$ & - & - & - & 14.5 & - & - & -\\
 \hline $\eta_b\to B_c$ & $f_{+(0)}(0)$ & 0.341[0.523] & - &
 - & - & - & - & - & - \\
& $f_+(q^2_{\rm max})$ & 0.976[0.918] & -
& - & - & - & - & - & - \\
& $f_0(q^2_{\rm max})$ & 0.811[0.839] & -
& - & - & - & - & - & - \\
& $\Gamma_{e(\mu)}$
& $ 4.64[7.94]$ & - & - & - & - & - & - & -\\
& $\Gamma_\tau$
& $1.57[2.11]$ & - & - & - & - & - & - & -\\
\hline\hline
\end{tabular}
\end{table*}

In Table~\ref{t3}, we summarize our results for the weak form
factors $f_+$ and $f_0$ at $q^2=0$ and $q^2_{\rm max}$ and the
decay widths $\Gamma_\ell$ of the semileptonic $B_c\to(D,\eta_c, B,
B_s)\ell\nu_\ell$ and $\eta_b\to B_c \ell\nu_\ell$($\ell=e,\mu,\tau$) decays
in comparison with other theoretical model
predictions\cite{IKS01,EFG03D,EFG03E,AMV,CD00,NW,HNV,WSL,AKN}. The
subscript for the decay width $\Gamma_{\ell}$ represents the
result for $P\to P\ell\nu_{\ell}$ decay where the final lepton is
$\ell=e,\mu$ or $\tau$. For the decays induced by $b\to u(c)$
transitions such as $B_c\to D,\eta_c$ and $\eta_b\to B_c$ decays,
we take $\Gamma_e\simeq\Gamma_\mu$ with the massless lepton limit
since the muon mass effect is negligible for these transitions
with large kinematic ranges. For the decays induced by $c\to d(s)$
transitions such as $B_c\to B(B_s)$ decays, $\Gamma_\mu$ is about
$5\%$ smaller than $\Gamma_e$ in our model predictions. For the
$B_c\to D$ decay, our predictions of the form factor $f_+$ at the
maximum-recoil point are rather smaller than other quark model
predictions. The upcoming experimental study planned at the Tevatron
and at the LHC may distinguish these different model predictions.
For the $B_c\to \eta_c, B$ and $B_s$ semileptonic decays, our predictions
are quite comparable with those of the quasipotential approach to
the relativistic quark model~\cite{EFG03D,EFG03E}, the
relativistic quark-meson model~\cite{NW}, and the nonrelativistic
quark model~\cite{HNV}. It may be noted, however, that the predictions
of the quark model based on an effective Lagrangian describing the
coupling of hadrons to their constituent quarks~\cite{IKS01} as
well as the covariant LFQM~\cite{WSL} are quite different from other model
predictions including ours.

Finally, in order to analyze the total rate for the radiative
$B^*_c\to B_c+\gamma$ decay, the masses of the $B_c$ and $B^*_c$
mesons must be specified. Although we predicted the above two
meson masses in Fig.~\ref{fig1}, we use the the central value of
the experimental data $M^{\rm exp}_{B_c}=6.276$ GeV~\cite{Data08}
to reduce the possible theoretical uncertainties. For the
unmeasured  $B^*_c$ meson mass, we take some range of the $B^*_c$
meson mass, i.e., $10\;{\rm MeV}\leq\Delta
m(=M_{B^*_c}-M_{B_c})\leq 220\; {\rm MeV}$. The upper value of
$\Delta m$(i.e., $M_{B^*_c}=6496$ MeV) is chosen to be
corresponding to our predictions, $M_{B^*_c}=6494$ MeV and 6492
MeV, obtained from the linear and HO potential models,
respectively.

\begin{figure}
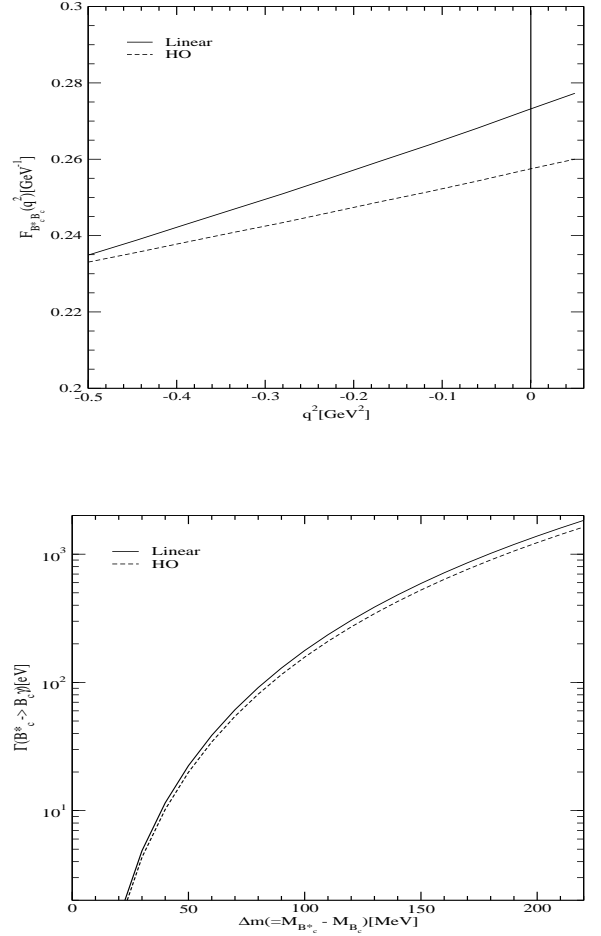

\vspace{1cm}
\includegraphics[width=3in,height=2.2in]{fig12a.eps}\\
\vspace{1.2cm}
\includegraphics[width=3in,height=2.2in]{fig12b.eps}
\caption{The transition form factor $F_{B^*_c B_c}(q^2)$(upper
panel) for $B^*_c\to B_c\gamma^*$ and the dependence of
$\Gamma(B^*_c\to B_c\gamma)$(lower panel) on $\Delta
m=M_{B^*_c}-M_{B_c}$ obtained from the linear and HO potential
parameters, where the central value of the experimental data
$M^{\rm exp}_{B_c}=6.276$ GeV~\cite{Data08} is used.}
\label{fig12}
\end{figure}

In Fig.~\ref{fig12}, we show the momentum-dependent form factor
$F_{B^*_c B_c}(q^2)$ (upper panel) for the radiative $B^*_c\to
B_c\gamma^*$ decay and the dependence of
$\Gamma(B^*_c\to B_c\gamma)$ on $\Delta m$ (lower panel) obtained from the
linear (solid line) and HO potential (dashed line) parameters. For
the transition form factor $F_{B^*_c B_c}(q^2)$, we have performed
the analytic continuation of $F_{B^*_c B_c}(q^2)$ from the
spacelike region ($q^2<0$) to the physical timelike $0\leq q^2\leq
q^2_{\rm max}$, where $q^2_{\rm max}=(M_{B^*_c}-M_{B_c})^2$
represents the zero recoil point of the $B_c$ meson. The coupling
constant $g_{B^*_c B_c}$ is obtained at the $q^2=0$ point that corresponds
to the $B_c$ meson recoiling with the maximum three-momentum in the rest
frame of the
$B^*_c$ meson. In our model calculation, the coupling constant
itself is independent of the physical masses of the mesons and our
prediction is $g_{B^*_c B_c}= 0.273\;[0.257]$ GeV$^{-1}$ for the
linear [HO] potential model. Our predictions are quite comparable
with the result from the QCD sum rule approach~\cite{AIP}, $g^{SR}_{B^*_c
B_c}=0.270\pm0.095$ GeV$^{-1}$. As one can see from the lower
panel of Fig.~\ref{fig12}, the dependence of $\Gamma(B^*_c\to
B_c\gamma)$ on $\Delta m$ is quite sensitive to the mass of the
$B^*_c$ meson, e.g., our linear [HO] potential model predicts
$\Gamma(B^*_c\to B_c\gamma)=22.4\;[19.9]\;{\rm eV}\sim
1836\;[1631]\;{\rm eV}$ for $\Delta m = 50\;{\rm MeV}\sim 220\;{\rm
MeV}$. This sensitivity for the $B^*_c$ radiative decay may help
in determining the mass of $B^*_c$ experimentally and pinning
down the best phenomenological model. Other magnetic dipole decays $V\to
P\gamma$ of various heavy-flavored mesons such as
$(D,D^*,D_s,D^*_s,\eta_c,J/\psi)$ and
$(B,B^*,B_s,B^*_s,\eta_b,\Upsilon)$ using our LFQM can be found
in~\cite{Choi07}.

\section{Summary and Discussion}
In this work, we investigated the exclusive semileptonic $B_c\to
(D,\eta_c,B,B_s)\ell\nu_\ell$, $\eta_b\to
B_c\ell\nu_\ell$ ($\ell=e,\mu,\tau$) decays and the magnetic dipole
$B^*_c\to B_c\gamma$ decay using our LFQM constrained by the
variational principle for the QCD motivated effective Hamiltonian
with the linear (or HO) plus Coulomb interaction. Especially, we
obtained the new variational parameter $\beta_{cb}$ for the
bottom-charm sector and predicted the mass eigenvalues of the
low-lying $B_c$ and $B^*_c$ states. Our new predictions of
$M_{B_c}=6459\;[6351]$ MeV obtained from the linear [HO] potential
model is in agreement with the data, $M^{\rm exp}_{B_c}=(6276\pm
4)$ MeV~\cite{Data08}, within 3$\%$ error. We also predicted the
unmeasured mass of $B^*_c$ as $M_{B^*_c}=6494\;[6496]$ MeV for the
linear [HO] potential model. Our model parameters obtained from the
variational principle uniquely determine the physical quantities
related to the above processes. This approach can establish the
broader applicability of our LFQM to the wider range of hadronic
phenomena. For instance, our LFQM has been tested extensively in
the spacelike processes~\cite{CJ1,CJKGPD} as well as in the
timelike exclusive processes such as
semileptonic~\cite{CJ2,JC,CJK} and rare~\cite{CJK02} decays of
pseudoscalar mesons and the magnetic dipole $V\to P\gamma^*$
decays~\cite{Choi07,Choi08}.

The weak form factor $f_{\pm}(q^2)$ for the semileptonic decays
between two pseudoscalar mesons and the decay form factor
$F_{B^*_cB_c}(q^2)$ for the $B^*_c\to B_c\gamma$ decay are
obtained in the $q^+=0$ frame ($q^2=-{\bf q}^2_\perp<0$) and then
analytically continued to the timelike region by changing ${\bf
q}^2_\perp$ to $-q^2$ in the form factor. The covariance (i.e.,
frame independence) of our model has been checked by performing the
LF calculation in the $q^+=0$ frame in parallel with the manifestly
covariant calculation using the exactly solvable covariant fermion
field theory model in $(3+1)$-dimensions. We found the zero-mode contribution
to the form factor $f_-(q^2)$ and identified the zero-mode operator that
is convoluted with the initial and final state LF wave functions.
We calculated the decay constants of $(B_c, B^*_c)$ mesons and the decay
rates for the exclusive $B_c\to (D,\eta_c,B,B_s)\ell\nu_\ell$ and
$\eta_b\to B_c\ell\nu_\ell$ decays and compared
with other theoretical approaches. Particularly, the decay
constants for $(B_c, B^*_c)$ mesons and the decay rate for $\eta_b\to B_c$ process
are quite sensitive to the choice of potential within our LFQM.
From the future experimental data on these sensitive processes, one may obtain
more realistic information on the potential between quark and antiquark in the
heavy meson system.

For the radiative $B^*_c\to
B_c\gamma$ decay, we find that the decay width $\Gamma(B^*_c\to
B_c\gamma)$ is very sensitive to the value of $\Delta
m=M_{B^*_c}-M_{B_c}$. This sensitivity for the $B^*_c$ radiative
decay may help in determining the mass of $B^*_c$ experimentally.
Since the form factor $F_{B^*_cB_c}(q^2)$ for the radiative
$B^*_c\to B_c\gamma$ decay presented in this work is
analogous to the vector current form factor $g(q^2)$ in the weak decay
of ground state vector meson to ground state pseudoscalar meson,
the ability of our model in describing the radiative decay
would therefore be relevant to the applicability of our model
also for the weak decay.
Consideration on such exclusive weak decays in our LFQM is
underway.

\acknowledgments The work of H.-M.Choi was supported by the Korea
Research Foundation Grant funded by the Korean
Government(KRF-2008-521-C00077) and that of C.-R.Ji by the U.S.
Department of Energy(No. DE-FG02-03ER41260).


\begin{thebibliography}{99}
\bibitem{Gouz} I. P. Gouz, V. V. Kiselev, A. K. Likhoded, V. I.
Romanovsky, and O. P. Yushchenko, Phys. Atom. Nucl. {\bf 67}, 1559
(2004); Yad. Fiz. {\bf 67}, 1581 (2004).
\bibitem{CNP} P. Colangelo, G. Nardulli, and N. Paver, \Journal{\ZPC}{57}{43}{1993}.
\bibitem{KKL} V. V. Kiselev, A. E. Kovalsky, and A. K. Likhoded,
\Journal{\NPB}{585}{353}{2000}; V. V. Kiselev, A. K. Likhoded, and
A. I. Onishchenko, \Journal{\NPB}{569}{473}{2000}.
\bibitem{HZ} T. Huang and F. Zuo, \Journal{\EPJC}{51}{833}{2007}.
\bibitem{IKS01} M. A. Ivanov, J. G. K\"{o}rner and P. Santorelli,
\Journal{\PRD}{63}{074010}{2001}.
\bibitem{IKS05} M. A. Ivanov, J. G. K\"{o}rner and P. Santorelli,
\Journal{\PRD}{71}{094006}{2005}.
\bibitem{IKS06} M. A. Ivanov, J. G. K\"{o}rner and P. Santorelli,
\Journal{\PRD}{73}{054024}{2006}.
\bibitem{EFG67} D. Ebert, R. N. Faustov and V. O. Galkin,
\Journal{\PRD}{67}{014027}{2003}.
\bibitem{EFG03D} D. Ebert, R. N. Faustov and V. O. Galkin, \Journal{\PRD}{68}{094020}{2003}.
\bibitem{EFG03E} D. Ebert, R. N. Faustov and V. O. Galkin, \Journal{\EPJC}{32}{29}{2003}.
\bibitem{CC94} C.-H. Chang and Y.-Q. Chen,
\Journal{\PRD}{49}{3399}{1994}.
\bibitem{LC97} J.-F. Liu and K.-T. Chao, \Journal{\PRD}{56}{4133}{1997}.
\bibitem{AMV} A. Abd El-Hady, J. H. Munoz, and J. P. Vary,
\Journal{\PRD}{62}{014019}{2000}.
\bibitem{CD00} P. Colangelo and F. De Fazio, \Journal{\PRD}{61}{034012}{2000}.
\bibitem{NW} M. A. Nobes and R. M. Woloshyn,
\Journal{\JPG}{26}{1079}{2000}.
\bibitem{HNV} E. Hern\'{a}ndez, J. Nieves and J. M. Verde-Velasco,
\Journal{\PRD}{74}{074008}{2006}.
\bibitem{WSL} W. Wang, Y.-L. Shen, and C.-D. L\"{u},
arXiv:0811.3748[hep-ph].
\bibitem{LM} M. Lusignoli and
M.Masetti,\Journal{\ZPC}{51}{549}{1991}.
\bibitem{DW}D. Du and Z. Wang, \Journal{\PRD}{39}{1342}{1989}.
\bibitem{DSV}R. Dhir, N. Sharma, and R.C.
Verma,\Journal{\JPG}{35}{085002}{2008}.
\bibitem{God} S. Godfrey, \Journal{\PRD}{70}{054017}{2004}.
\bibitem{BSW} M. Wirbel, B. Stech, and M. Bauer,
\Journal{\ZPC}{29}{637}{1985}; M. Bauer, B. Stech, and M. Wirbel,
\Journal{\ZPC}{34}{103}{1987}.
\bibitem{ISGW}N. Isgur, D. Scora, B. Grinstein, and
M.B. Wise, \Journal{\PRD}{39}{799}{1989}.
\bibitem{CJ1}H.-M. Choi and C.-R.
Ji,\Journal{\PRD}{59}{074015}{1999}.
\bibitem{CJ2}H.-M. Choi and C.-R. Ji,\Journal{\PLB}{460}{461}{1999}.
\bibitem{JC} C.-R. Ji and H.-M. Choi, \Journal{\PLB}{513}{330}{2001}.
\bibitem{CJK02}H.-M. Choi, C.-R. Ji, and
L.S. Kisslinger, \Journal{\PRD}{65}{074032}{2002}.
\bibitem{Choi07} H.-M. Choi, \Journal{\PRD}{75}{073016}{2007};
J. Korean Phys. Soc. {\bf 53}, 1205 (2008).
\bibitem{Choi08} H.-M. Choi, \Journal{\PRD}{77}{097301}{2008}.
\bibitem{Zero} H.-M. Choi and C.-R. Ji,
\Journal{\PRD}{58}{071901(R)}{1998};
\Journal{\PRD}{72}{013004}{2005}; S. J. Brodsky and D. S. Hwang,
\Journal{\NPB}{543}{239}{1998}; M. Burkardt, \Journal{\PRD}{47}{4628}{1993};
J.P.B.C. de Melo, J.H.O. Sales, T. Frederico, and P.U. Sauer,
\Journal{\NPA}{631}{574}{1998c}
\bibitem{Jaus99} W. Jaus, \Journal{\PRD}{60}{054026}{1999}.
\bibitem{CJK} H.-M. Choi and C.-R. Ji,
\Journal{\PRD}{59}{034001}{1998}.
\bibitem{BCJ01} B.L.G. Bakker, H.-M. Choi, and C.-R. Ji,
 \Journal{\PRD}{63}{074014}{2001}.
\bibitem{BCJ03} B.L.G. Bakker, H.-M. Choi, and C.-R. Ji,
 \Journal{\PRD}{67}{113007}{2003}.
 \bibitem{MF} J.P.B.C. de Melo and T. Frederico,
 \Journal{\PRC}{55}{2043}{1997}; J.P.B.C. de Melo, T. Frederico, E. Pace, and
 G. Salme, \Journal{\PRD}{73}{074013}{2006}.
\bibitem{CDKM} J. Carbonell, B. Desplanques, V.A. Karmanov, and J.-F. Mathiot,
Phys. Rep. {\bf 300}, 215 (1998).
\bibitem{SCC} A. Szczepaniak, C.-R. Ji, and S.R. Cotanch, \Journal{\PRD}{52}{5284}{1995}.
\bibitem{CJ08D} H.-M. Choi and C.-R. Ji,\Journal{\PRD}{75}{034019}{2007}.
\bibitem{SI} D. Scora and N. Isgur, \Journal{\PRD}{52}{2783}{1995}.
\bibitem{Data08} C. Amsler {\em et al.}(Particle Data Group),
\Journal{\PLB}{667}{1}{2008}.
\bibitem{BaBar08} P. Grenier, arxiV:0809.1672 [hep-ex].
\bibitem{EQ} E. J. Eichten and C. Quigg, \Journal{\PRD}{49}{5845}{1994}.
\bibitem{GKLT} S. S. Gershtein, V. V. Kiselev, A. K. Likhoded, and
A.V. Tkabladze, \Journal{\PRD}{51}{3613}{1995}.
\bibitem{Ful}L. P. Fulcher, \Journal{\PRD}{60}{074006}{1999}.
\bibitem{CG} S. Capstick and S. Godfrey, \Journal{\PRD}{41}{2856}{1990}.
\bibitem{LB} G. P. Lepage and S. J. Brodsky, \Journal{\PRD}{22}{2157}{1980}.
 \bibitem{JLMS} E. Jenkins, M. Luke, A. V. Manohor, and M. J.
Savage, \Journal{\NPB}{390}{463}{1993}.
\bibitem{AKN} A.Yu. Anisimov, P.
Yu. Kulikov, I.M. Narodetskii, and K.A. Ter-Martirosyan, Phys.
Atom Nucl. {\bf 62}, 1739 (1999)[Yad. Fiz. {\bf 62}, 1868 (1999)].
\bibitem{AIP} T. M. Aliev, E. Iltan, and N. K. Pak,
\Journal{\PLB}{329}{123}{1994}.
\bibitem{CJKGPD} H.-M. Choi, C.-R. Ji, and
L.S. Kisslinger, \Journal{\PRD}{64}{093006}{2001};
\Journal{\PRD}{66}{053011}{2002}.
\end{thebibliography}
\end{document}